\documentclass[12pt,preprint]{aastex}

\pdfoutput=1

\usepackage{color,hyperref}
\definecolor{linkcolor}{rgb}{0,0,0.5}
\hypersetup{colorlinks=true,linkcolor=linkcolor,citecolor=linkcolor,
            filecolor=linkcolor,urlcolor=linkcolor}
\usepackage{url}
\usepackage{amssymb,amsmath}
\usepackage{subfigure}
\usepackage{booktabs}

\usepackage{natbib}
\newcommand{\project}[1]{\textsl{#1}} % hogg say
\newcommand{\kepler}{\project{Kepler}}
\newcommand{\KT}{\project{K2}}
\newcommand{\tess}{\project{TESS}}
\newcommand{\jwst}{\project{JWST}}

\newcommand{\pdc}{\project{PDC}}

\newcommand{\paper}{\textsl{Article}}

\newcommand{\foreign}[1]{\emph{#1}}
\newcommand{\etal}{\foreign{et\,al.}}

\newcommand{\figref}[1]{\ref{fig:#1}}
\newcommand{\Fig}[1]{\figurename~\figref{#1}}
\newcommand{\fig}[1]{\Fig{#1}}
\newcommand{\figlabel}[1]{\label{fig:#1}}
\newcommand{\Tab}[1]{Table~\ref{tab:#1}}
\newcommand{\tab}[1]{\Tab{#1}}
\newcommand{\tablabel}[1]{\label{tab:#1}}
\newcommand{\Eq}[1]{Equation~(\ref{eq:#1})}
\newcommand{\eq}[1]{\Eq{#1}}
\newcommand{\eqalt}[1]{Equation~\ref{eq:#1}}
\newcommand{\eqlabel}[1]{\label{eq:#1}}
\newcommand{\sectionname}{Section}
\newcommand{\Sect}[1]{\sectionname~\ref{sect:#1}}
\newcommand{\sect}[1]{\Sect{#1}}

\newcommand{\App}[1]{Appendix~\ref{sect:#1}}
\newcommand{\app}[1]{\App{#1}}
\newcommand{\sectlabel}[1]{\label{sect:#1}}

\newcommand{\BIC}{{\ensuremath{\mathrm{BIC}}}}
\newcommand{\T}{\ensuremath{\mathrm{T}}}

\newcommand{\bvec}[1]{{\ensuremath{\boldsymbol{#1}}}}
\newcommand{\appropto}{\mathrel{\vcenter{
  \offinterlineskip\halign{\hfil$##$\cr
    \propto\cr\noalign{\kern2pt}\sim\cr\noalign{\kern-2pt}}}}}

\definecolor{mygreen}{rgb}{0, 0.50196, 0}
\newcommand{\response}[1]{#1}
\newcommand{\flux}{{\ensuremath{\bvec{f}}}}

\newcommand{\period}{{\ensuremath{P}}}
\newcommand{\phase}{{\ensuremath{T^0}}}
\newcommand{\duration}{{\ensuremath{D}}}
\newcommand{\depth}{{\ensuremath{Z}}}
\newcommand{\transittime}{{\ensuremath{T}}}
\newcommand{\impact}{{\ensuremath{b}}}
\newcommand{\ecc}{{\ensuremath{e}}}
\newcommand{\pomega}{{\ensuremath{\omega}}}

\newcommand{\datareleaseurl}{{\url{http://bbq.dfm.io/ketu}}}

\begin{document}

\title{%
    A systematic search for transiting planets in the \KT\ data
}

\newcommand{\nyu}{2}
\newcommand{\caltech}{3}
\newcommand{\cfa}{4}
\newcommand{\mpia}{5}
\newcommand{\cds}{6}
\newcommand{\princeton}{7}
\newcommand{\mpis}{8}
\author{%
    Daniel~Foreman-Mackey\altaffilmark{1,\nyu},
    Benjamin~T.~Montet\altaffilmark{\caltech,\cfa},
    David~W.~Hogg\altaffilmark{\nyu,\mpia,\cds},
    Timothy~D.~Morton\altaffilmark{\princeton},
    Dun~Wang\altaffilmark{\nyu}, \&
    Bernhard~Sch\"olkopf\altaffilmark{\mpis}
}
\altaffiltext{1}         {To whom correspondence should be addressed:
                          \url{danfm@nyu.edu}}
\altaffiltext{\nyu}      {Center for Cosmology and Particle Physics,
                          Department of Physics, New York University,
                          4 Washington Place, New York, NY, 10003, USA}
\altaffiltext{\caltech}  {Cahill Center for Astronomy and Astrophysics,
                          California Institute of Technology, Pasadena, CA,
                          91125, USA}
\altaffiltext{\cfa}      {Harvard-Smithsonian Center for Astrophysics,
                          Cambridge, MA 02138, USA}
\altaffiltext{\mpia}     {Max-Planck-Institut f\"ur Astronomie,
                          K\"onigstuhl 17, D-69117 Heidelberg, Germany}
\altaffiltext{\cds}      {Center for Data Science,
                          New York University,
                          726 Broadway, 7th Floor, New York, NY, 10003, USA}
\altaffiltext{\princeton}{Department of Astrophysics, Princeton University,
                          Princeton, NJ, 08544, USA}
\altaffiltext{\mpis}     {Max Planck Institute for Intelligent Systems,
                          Spemannstrasse 38, 72076 T\"ubingen, Germany}

\begin{abstract}

Photometry of stars from the \KT\ extension of NASA's \kepler\ mission is
afflicted by systematic effects caused by small (few-pixel) drifts in the
telescope pointing and other spacecraft issues.
We present a method for searching \KT\ light curves for evidence of
exoplanets by simultaneously fitting for these systematics and the
transit signals of interest.
This method is more computationally expensive than standard search algorithms
but we demonstrate that it can be efficiently implemented and used to
discover transit signals.
We apply this method to the full Campaign~1 dataset and report a list of 36
planet candidates transiting 31 stars, along with an analysis of the pipeline
performance and detection efficiency based on artificial signal injections
and recoveries.
For all planet candidates, we present posterior distributions on the
properties of each system based strictly on the transit observables.

\end{abstract}


\section{Introduction}

The \kepler\ Mission was incredibly successful at finding transiting
exoplanets in the light curves of stars.
The Mission has demonstrated that it is possible to routinely measure signals
in stellar light curves at the part-in-$10^5$ level.
Results from the primary mission include the detection of planet transits with
depths as small as 12 parts per million \citep{Barclay:2013}.

The noise floor for \kepler\ data is often quoted as 15 parts per million
(ppm) per six hours of observations \citep{Gilliland:2011}.
Although they generally do not interfere with searches for transiting
planets, larger systematic effects exist on different timescales.
One of the most serious of these is spacecraft pointing: If the detector
flat-field is not known with very high accuracy, then tiny changes to the
relative illumination of pixels caused by a star's motion in the focal plane
will lead to changes in the measured or inferred brightness of the star.

The great stability of the original \kepler\ Mission
came to an end with the failure of a critical reaction wheel.
The \KT\ Mission \citep{Howell:2014} is a follow-on to the primary Mission,
observing about a dozen fields near the ecliptic plane, each for
$\sim 75$ days at a time.
Because of the degraded spacecraft orientation systems, the new \KT\ data
exhibit far greater pointing variations---and substantially more
pointing-induced variations in photometry---than the original \kepler\ Mission
data.
This makes good data-analysis techniques even more valuable.

Good photometry relies on either a near-perfect flat-field
and pointing model or else data-analysis techniques that are
insensitive to these instrument properties.
The flat-field for \kepler\ was measured on the ground before the launch of
the spacecraft, but is not nearly as accurate as required to make
pointing-insensitive photometric measurements at the relevant level of
precision.
In principle direct inference of the flat-field might be possible;
however, because point sources are observed with relatively limited
spacecraft motion, and only a few percent of the data are actually stored and
downloaded to Earth, there isn't enough information in the data to derive or
infer a complete or accurate flat-field map.
Therefore, work on \KT\ is sensibly focused on building data-analysis
techniques that are pointing-insensitive.

Previous projects have developed methods to work with \KT\ data.
Both \citet{Vanderburg:2014} and \citet{Armstrong:2014}
extract aperture photometry from the pixel data
and decorrelate with image centroid position, producing light curves for each
star that are ``corrected'' for the spacecraft motion.
These data have produced the first confirmed planet found with
\KT\ \citep{Vanderburg:2015}.
Both \citet{Aigrain:2015} and \citet{Crossfield:2015} use a Gaussian Process
model for the measured flux, with pointing measurements as the inputs, and
then ``de-trend'' using the mean prediction from that model.
Other data-driven approaches have been developed and applied to the data from
space missions \citep[for example,][]{Ofir:2010, Stumpe:2012, Smith:2012,
Petigura:2013, Wang:2015} and ground-based surveys \citep[for
example,][]{Kovacs:2005, Tamuz:2005, Berta:2012} but they have yet to be
generalized to \KT.

In all of these light-curve processing methodologies, the authors follow a
traditional procedure of ``correcting'' or ``de-trending'' the light curve to
remove systematic and stellar variability as a step that happens \emph{before}
the search for transiting planets.
Fit-and-subtract is dangerous:
Small signals, such as planet transits, can be
partially absorbed into the best-fit stellar variability or systematics
models, making each individual transit event appear shallower.
In other words, the traditional methods are prone to over-fitting.
Because over-fitting will in general reduce the amplitude of true exoplanet
signals, small planets that ought to appear just above any specific
signal-to-noise or depth threshold could be missed because of the de-trending.
This becomes especially important as the amplitude of the noise increases.

The alternative to this approach is to \emph{simultaneously fit} both the
systematics and the transit signals.
Simultaneous fitting can push the detection limits to lower signal-to-noise
while robustly accounting for uncertainties about the systematic trends.
In particular, it permits us to \emph{marginalize} over choices in the noise
model and propagate any uncertainties about the systematic effects
to our confidence in the detection.
This marginalization ensures that any conclusions we come to about the
exoplanet properties are conservative, given the freedom of the systematics
model.

In this \paper\ we present a data-analysis technique for exoplanet search and
characterization that is insensitive to spacecraft-induced trends in the light
curves.
We assume that the dominant trends in the observed light curves in each star
are caused by the spacecraft and are, therefore, shared with other stars.
We reduce the dimensionality by running PCA on stellar light curves to obtain
the dominant modes.
The search for planets proceeds by modeling the data as a linear combination
of 150 of these basis vectors and a transit model.
Our method builds on the ideas behind previous data-driven de-trending
procedures such as the \kepler\ pipeline pre-search data conditioning
\citep[\pdc;][]{Stumpe:2012, Smith:2012}, but (because of our simultaneous
fitting approach) we can use a much more flexible systematics model
\response{while being less prone to over-fitting}.

The methods developed within this paper are highly relevant to both \KT\ and
the upcoming \tess\ mission \citep{Ricker:2014}.
\tess\ will feature pointing precision of $\sim
3$~arcseconds\footnote{\url{http://tess.gsfc.nasa.gov/documents/TESS_FactSheet_Oct2014.pdf}},
similar to the level of pointing drift with \KT.
Moreover, the typical star will be only observed for one month at a time, and
the typical transit detection will be at a similar signal-to-noise ratio as
with \KT.

Catalogs of transiting planets found in the \KT\ data will be important to
better understand the physical properties, formation, and evolution of
planetary systems.
These planets, especially when they orbit bright or late-type stars, will be
useful targets for ground-based and space-based follow-up, both for current
facilities and those planned in the near future such as \jwst.
They will also deliver input data for next-generation population inferences
\citep{Foreman-Mackey:2014}, especially for the population of planets around
cool stars \citep[for example,][]{Dressing:2015}.

This project follows in the tradition of independently implemented transit
search algorithms applied to publicly available datasets \citep[such
as][]{Petigura:2013a, Petigura:2013, Sanchis-Ojeda:2014, Dressing:2015}.
These efforts have been hugely successful, especially in the field of
exoplanet population inference because, thanks to their relative simplicity,
the efficiency and behavior of these pipelines can be quantified empirically.
The work described in this \paper\ is built on many of the same principles as
the previous projects developed for studying \kepler\ data but our main
intellectual contribution is a computationally tractable framework for
simultaneously fitting for the trends and the transit signal even when
searching for planets.

The \paper\ is organized as follows.
In \sect{phot}, we describe our method of extracting aperture photometry from
the calibrated \KT\ postage stamp time series.
In \sect{model} (with details in \app{math}), we describe our data-driven
model for the systematic trends in the photometric light curves and our method
for fitting this model simultaneously with a transit signal.
In \sect{search}, we give the detailed procedure that we use for discovering
and vetting planet candidates.
To quantify the performance and detection efficiency of our pipeline, we test
(in \sect{perform}) the recovery of synthetic transit
signals, spanning a large range of physical parameters, injected into real
\KT\ light curves.
Finally, in \sect{results}, we present a catalog of 36 planet candidates
orbiting 31 stars from the publicly available \KT\ Campaign~1 dataset.

\section{Photometry and Eigen Light Curves}
\sectlabel{phot}

The starting point for analysis is the raw pixel data.
We download the full set of 21,703 target pixel files for \KT's Campaign~1
from MAST\footnote{\url{https://archive.stsci.edu/k2/}}.
We extract photometry using fixed, approximately circular, binary apertures of
varying sizes centered on the predicted location of the target star based on
the world coordinate system.
For each target, we use a set of apertures ranging in radius from 1 to 5
pixels (in steps of 0.5 pixels).
Following \citet{Vanderburg:2014}, we choose the aperture size with the
minimum CDPP \citep{Christiansen:2012} with a 6 hour window.\footnote{Note
that although we chose a specific aperture for each star, photometry for every
aperture radius is available online: \datareleaseurl.}

All previous methods for analyzing \KT\ data involve some sort of
``correction'' or ``de-trending'' step based on measurements of the pointing
of the spacecraft \citep{Vanderburg:2014, Aigrain:2015, Crossfield:2015}.
In our analysis, we do not do any further preprocessing of the light curves
because, as we describe in the next \sectionname, we fit raw photometric light
curves with a model that includes both the trends and the transit signal.

One key realization that is also exploited by the official \kepler\ pipeline
is that the systematic trends caused by pointing shifts and other instrumental
effects are shared---with different signs and weights---by all the stars on
the focal plane.
For a rigorous theoretical analysis of this problem, see
\citep{Scholkopf:2015}.
To capitalize on this, the \pdc\ component of the \kepler\ pipeline
removes any trends from the light curves that can be fit using a linear
combination of a small number of ``co-trending basis vectors''.
This basis of trends was found by running Principal Component Analysis (PCA)
on a large set of (filtered) light curves and extracting the top few ($\sim
4$) components \citep{Stumpe:2012, Smith:2012}.
Similarly, we ran PCA on the full set of our own generated \KT\ Campaign~1
light curves to determine a basis of representative trends but, unlike \pdc,
we retain and use a larger number of these components ($150$).
For clarity, we will refer to our basis as a set of ``eigen light curves''
(ELCs) and the full set is made available online\footnote{\datareleaseurl}.
The top ten ELCs for Campaign~1 are shown in \Fig{pca}.

\begin{figure}[p]
\begin{center}
\includegraphics{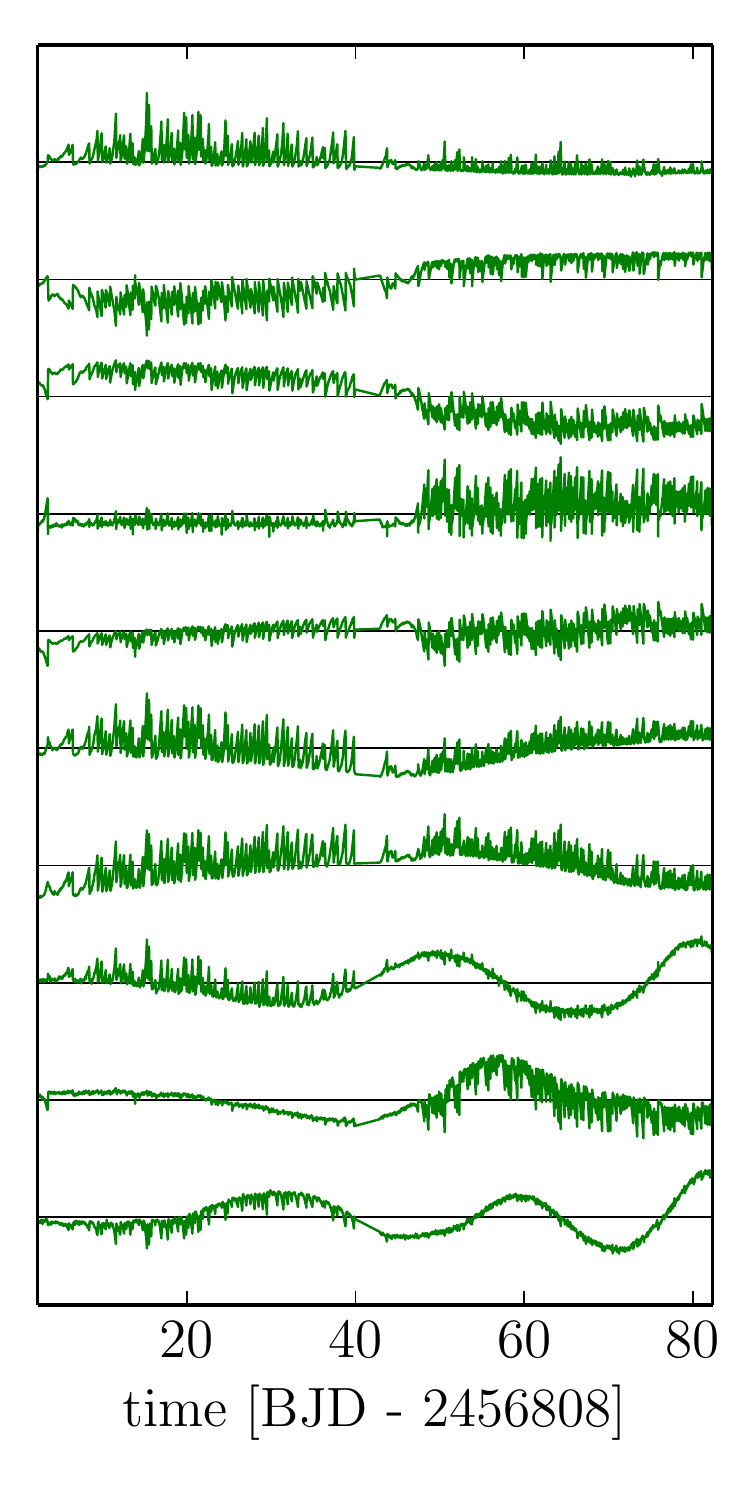}
\end{center}
\caption{%
The top 10 eigen light curves (ELCs) generated by running principal component
analysis on all the aperture photometry from Campaign~1.
\figlabel{pca}}
\end{figure}

\section{Joint transit \& variability model}
\sectlabel{model}

The key insight in our transit search method that sets it apart from most
standard procedures is that no de-trending is necessary.
Instead, we can fit for the noise (or trends) and exoplanet signals
simultaneously.
This is theoretically appealing because it should be more sensitive to low
signal-to-noise transits \response{and similar methods have been shown to be
effective for finding transits in ground-based surveys \citep{Berta:2012}}.
The main motivation for this model is that the signal is never precisely
orthogonal to the systematics and any de-trending will over-fit.
This will, in turn, decrease the amplitude of the signal and distort its
shape.
In order to reduce these effects, most de-trending procedures use a very rigid
model for the systematics.
For \KT, this rigidity has been implemented by effectively asserting that
centroid measurements contain all of the information needed to describe the
trends \citep{Vanderburg:2014, Aigrain:2015, Crossfield:2015}.
In the \kepler\ pipeline, this is implemented by allowing only a small number
of PCA components to contribute to the fit in the \pdc\ procedure.
Instead, we will use a large number of ELCs---a very flexible model---and use
a simultaneous fitting and marginalization to avoid over-fitting.

Physically, the motivation for our model---and the \pdc\ model---is that every
star on the detector should be affected by the same set of systematic effects.
These are caused by things like pointing jitter, temperature variations, and
other sources of PSF modulation.
Each of these effects will be imprinted in the light curves of many stars with
varying amplitudes and signs as a result of the varying flat field and PSF.
Therefore, while it is hard to write down a physical generative model for the
systematics, building a data-driven model might be possible.
This intuition is also exploited by other methods that model the systematics
using only empirical centroids \citep{Vanderburg:2014, Armstrong:2014,
Aigrain:2015, Crossfield:2015}, but our more flexible model should capture a
wider range of effects, including those related to PSF and temperature.
For example, \Fig{corr} shows the application of our model---with 150
ELCs---to a light curve with no known transit signals and the photometric
precision is excellent.

If we were to apply this systematics model alone (without a simultaneous fit
of the exoplanet transit model) to a light curve with transits, we would be at
risk of over-fitting and decreasing the amplitude of the signal.
\response{\Fig{overfitting} demonstrates this effect on a synthetic transit
injected into the light curve of a typical bright star.
The middle two panels in this Figure show the light curve de-trended using 10
and 150 ELCs respectively.
When only 10 ELCs are used, the measured transit depth is relatively robust
but this model is clearly not sufficient for removing the majority of the
systematic trends.
The model with 150 ELCs does an excellent job of removing the systematics but
it also distorts the transit shape and decreases the measured transit depth,
hence reducing the signal strength in the \project{BLS} spectrum
\citep{Kovacs:2002}.}

In our pipeline we simultaneously fit for the transit signal and the trends
using a rigid model for the signal and a relatively flexible model for the
systematic noise.
Specifically, we model the light curve as being generated by linear
combination of 150 ELCs and a ``box'' transit model at a specific period,
phase, and duration.
The mathematical details are given in \app{math}, but in summary, since the
model is linear, we can analytically compute the likelihood
function---conditioned on a specific period, phase, and duration---for the
depth \emph{marginalizing out the parameters of the systematics model}.
The signal-to-noise of this depth measurement can then be used as a quality of
fit metric or candidate selection scalar.
This computation is expensive but, as described in the following \sectionname
s, it is possible to scale the method to a \KT-size dataset.
\response{The bottom panel of \Fig{overfitting} shows the application of this
joint transit--systematics model to the synthetic transit discussed
previously.
When the joint model is used, the correct transit depth is measured---the
transit is not distorted---but the systematics are also well-described by the
model.}

It is worth noting that this model can be equivalently thought of as a
(computationally expensive) generalization of the ``Box Least Squares''
\citep[\project{BLS};][]{Kovacs:2002} method to a more sophisticated
description of the noise and systematics.
Therefore, any existing search pipeline based on \project{BLS} could, in
theory, use this model as a drop-in replacement, although some modifications
might be required for computational tractability.

\response{%
The choice to use 150 basis functions is largely arbitrary and we make no
claims of optimality.
This value was chosen as a trade-off between the computational cost of the
search---the cost scales as the third power of the size of the basis---and the
predictive power of the model.
In some preliminary experiments, we found that using a larger basis did, as
expected, lead to a marginally higher sensitivity to small transit signals
but the gain wasn't sufficient to justify the added cost.
}

\begin{figure}[p]
\begin{center}
\includegraphics{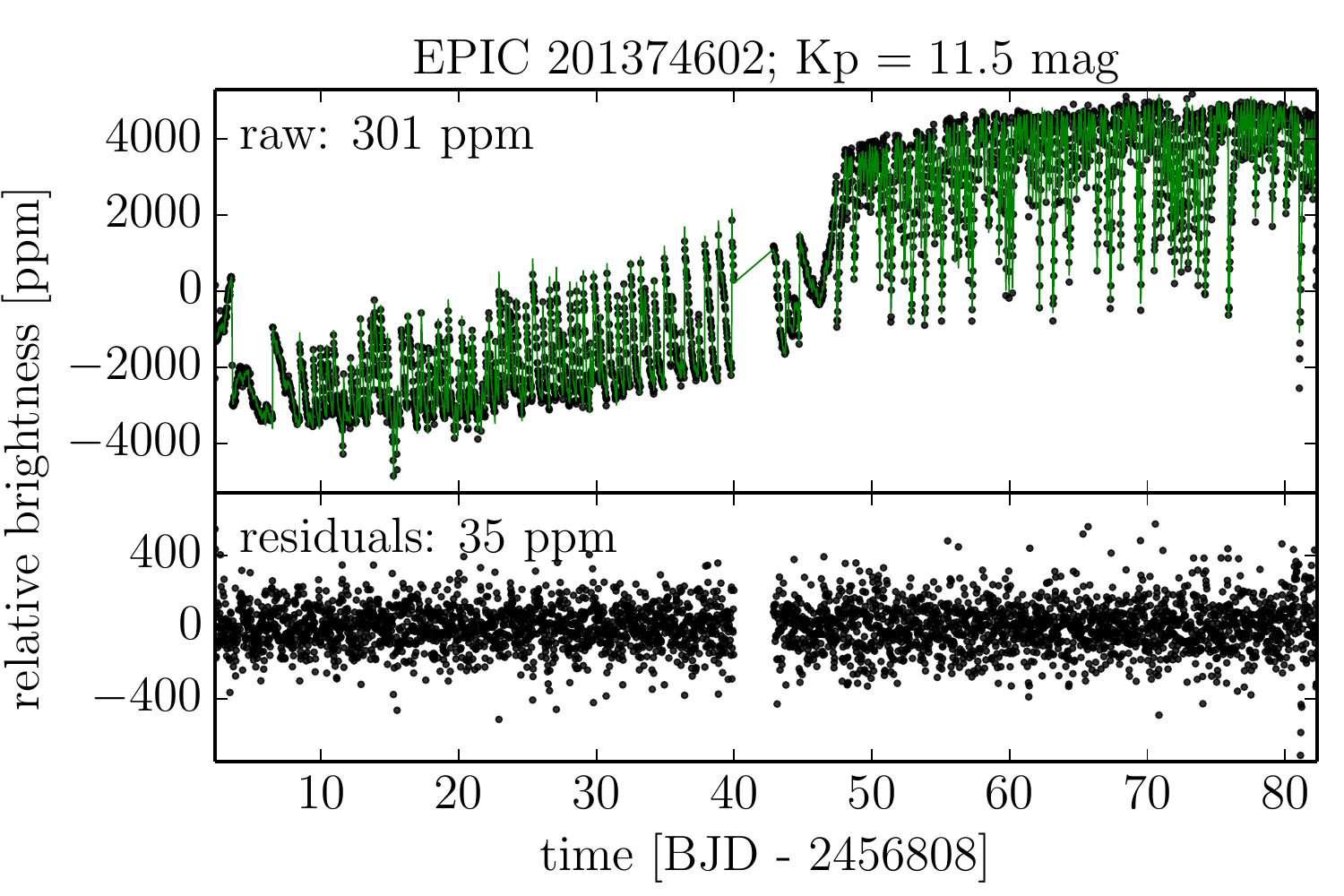}
\end{center}
\caption{%
A demonstration of the eigen light curve (ELC) fit to the aperture photometry
for EPIC 201374602.
\emph{Top:} The black points show the aperture photometry and the green line
is the maximum likelihood linear combination of ELCs.
The estimated 6-hour precision of the raw photometry is 264 ppm.
\emph{Bottom:} The points show the residuals of the data away from the ELC
prediction.
The 6-hour precision of this light curve is 31 ppm.
Note that although we show a ``de-trended'' light curve to give a qualitative
understanding of the model, this is not a product of the analysis.
In this search for transits, \emph{the data are only de-trended for the
purpose of visualization}.
\figlabel{corr}}
\end{figure}

\begin{figure}[p]
\begin{center}
\includegraphics[width=0.5\textwidth]{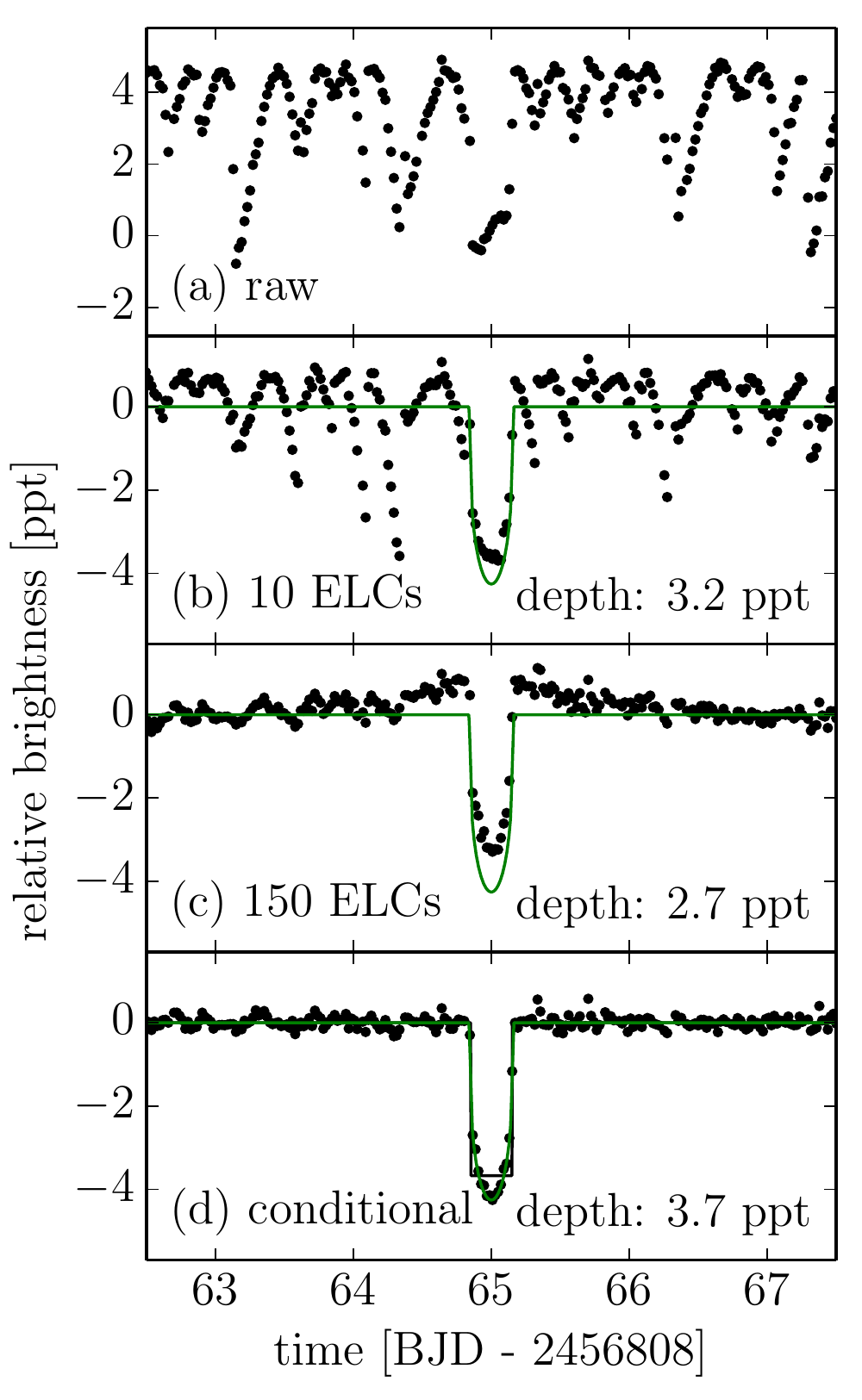}
\end{center}
\caption{%
\response{%
A comparison between de-trending using different numbers of ELCs and a
simultaneous fit of the systematics and transit model.
\emph{(a)} The raw photometry for EPIC 201374602 with a synthetic transit
injected at 65 days.
% In all plots, the photometry is measured in parts-per-thousand (ppt).
\emph{(b)} The black points show the photometry de-trended using a linear
combination of 10 ELCs and the green line shows the true transit model.
In this panel and the next, the maximum likelihood transit depth is computed
following \project{BLS} \citep{Kovacs:2002}.
While some of the systematics are removed by this model, there is still a lot
of residual noise.
\emph{(c)} The same plot as panel \emph{(b)} but using 150 ELCs to de-trend.
This model removes the majority of the systematics but also distorts the
transit and weakens the signal; it reduces the measured transit depth.
\emph{(d)} The final panel shows the results of simultaneously fitting for
the transit and the systematics using 150 ELCs.
The maximum likelihood depth (marginalized over the ELC weights) is computed
as described in \app{math}.
Like panel \emph{(c)}, this model removes most of the systematics but does
not distort the transit or reduce the measured transit depth.
}
\figlabel{overfitting}}
\end{figure}

\section{Search pipeline}
\sectlabel{search}

In principle, the search for transit signals simply requires evaluation of the
model described above on a fine three-dimensional grid in period, phase, and
duration, and then detection of high significance peaks in that space.
In practice, this is computationally intractable for any grids of the
required size and resolution.
Instead, we can compute the values on this grid approximately, but at very
high precision, using a two-step procedure that is much more efficient.

Specifically, we must evaluate the likelihood function for the light curve
$\flux_n$ of star $n$ given a period \period, reference transit time \phase,
duration \duration, \response{and depth \depth}
\begin{eqnarray}\eqlabel{full-likelihood}
p(\flux_n\,|\,\period,\,\phase,\,\duration,\,\depth) \quad.
\end{eqnarray}
We make the simplifying assumption that each transit enters this quantity
independently.
This is not true; as we change beliefs about each transit, we change beliefs
about the systematics model, which in turn affects the other transits.
However, this simplifying assumption
is approximately satisfied for all but the shortest
periods and leads to a huge computational advantage.
Under this assumption, this likelihood function can be rewritten as
\begin{eqnarray}\eqlabel{indtran}
p(\flux_n\,|\,\period,\,\phase,\,\duration,\,\depth) &=&
\prod_{m=1}^{M(\period,\,\phase)}
    p(\flux_n\,|\,\transittime_m(\period,\,\phase),\,\duration,\,
                    \depth)
\end{eqnarray}
where $\transittime_m(\period,\,\phase)$ is the time of the $m$-th
transit given the period $\period$ and reference time $\phase$, and
\response{$M(\period,\,\phase)$ is the total number of transits in the dataset
for the given \period\ and \phase}.
\Eq{indtran} can be efficiently computed for many periods and phases if we
first compute a set of likelihood functions for single transits on a grid
in $\transittime_l$ and duration $\duration_k$
\begin{eqnarray}\eqlabel{singletransit}
\left \{ p(\flux_n\,|\,\transittime_l,\,\duration_k,\,\depth)
\right\}_{l=1,\,k=1}^{L,\,K} \quad.
\end{eqnarray}
Then, we can use these results as a look-up table\response{---with
nearest-neighbor interpolation---}to approximately evaluate the full
likelihood in \eq{full-likelihood}.

In the remainder of this \sectionname, we give more details about each step of
the search procedure.
In summary, it breaks into three main steps: linear
search, periodic search, and vetting.
In the {\bf linear search} step, we evaluate the likelihood function in
\eq{singletransit} on a two-dimensional grid, coarse in transit duration
$\duration_k$ and fine in transit time $\transittime_m$.
Then in the {\bf periodic search} step, we use this two-dimensional grid to
approximately evaluate the likelihood (\eqalt{indtran}) for a
three-dimensional grid of periodic signals.
Then, we run a peak detection algorithm on this grid \response{that is robust
to signals with substantially varying transit depths}.
These transit candidates are then passed along for machine and human {\bf
vetting}.

\paragraph{Linear search}

The linear search requires hypothesizing a set of transit signals on a
two-dimensional grid in transit time and duration.
For each point in the grid, we use the model described in \sect{model} to
evaluate \emph{the likelihood function} for the transit depth at that time
and duration.
Since the model is linear and the uncertainties are assumed Gaussian, the
likelihood function for the depth (marginalized over the model of the
systematics) is a Gaussian with analytic amplitude $L$, mean $\bar{\depth}$,
and variance $\delta\bar{\depth}^2$, all derived and given in \app{math}.
In the linear search, we save these three numbers on a
two-dimensional grid of transit times $\transittime_l$ and durations
$\duration_k$.
The transit time grid spans the full length of Campaign~1 with half hour spacing
and we choose to only test three durations: 1.2, 2.4, and 4.8 hours.
\Fig{linear} shows the maximum likelihood transit depth $\bar{\depth}$ as a
function of transit time \transittime\ for the light curve of EPIC 201613023,
a transiting planet candidate with a period of 8.3~days.

\begin{figure}[p]
\begin{center}
\includegraphics{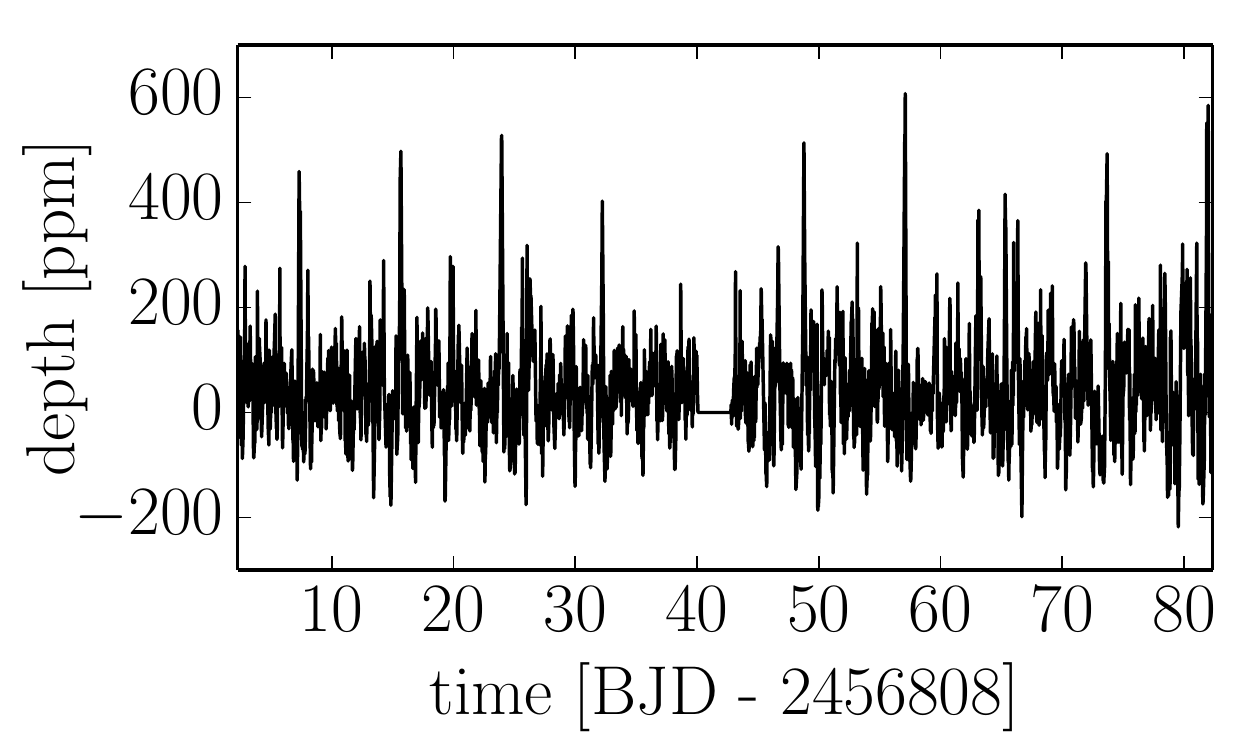}
\end{center}
\caption{%
The maximum likelihood transit depth as a function of transit time as computed
in the linear search of the light curve of EPIC 201613023.
After the periodic search and vetting this target is found to have a planet
candidate with a period of 8.3~days.
The first transit occurs at 7.4~days on this plot.
\figlabel{linear}}
\end{figure}

\paragraph{Periodic search}

\response{%
In the period search step, the table of likelihood functions generated in the
linear search step are used to compute the likelihood of the periodic model
(\eqalt{indtran}) on a three dimensional grid in period \period, reference
time \phase, and duration \duration.
At each point in this grid, the likelihood function for each transit depth is
chosen as the nearest point (without interpolation) calculated in the linear
search.
If the time spacing of the linear search is sufficiently fine, this will give
a good approximation of the correct periodic likelihood.}
For each periodic model, we compute the likelihood of a model where the
transit depth varies between transits and the ``correct'' simpler model where
the transit depth is constant.
The variable depth likelihood is given by the product of amplitudes from the
initial search
\begin{eqnarray}
p_\mathrm{var}(\flux_n\,|\,\period,\,\phase,\,\duration) &=&
\prod_{m=1}^{M(\period,\,\phase)} L_m \quad.
\end{eqnarray}
Since the likelihood function for the depth at each transit time is known and
Gaussian, the likelihood function for the depth under the periodic model can
also be computed analytically; it is a product of Gaussians which itself is a
Gaussian
\begin{eqnarray}
p_\mathrm{const}(\flux_n\,|\,\period,\,\phase,\,\duration) &=&
\prod_{m=1}^{M(\period,\,\phase)}
    \frac{L_m}{\sqrt{2\,\pi\,{\delta\bar{\depth}_m}^2}}\,\exp \left(
        -\frac{[\depth - \bar{\depth}_m]^2}{2\,{\delta\bar{\depth}_m}^2}
    \right)
\end{eqnarray}
where the maximum likelihood depth, for the periodic model, is
\begin{eqnarray}\eqlabel{periodic-depth}
\depth &=& {\sigma_\depth}^2\,\sum_{m=1}^{M(\period,\,\phase)}
    \frac{\bar{\depth}_m}{{\delta\bar{\depth}_m}^2}
\end{eqnarray}
and the uncertainty is given by
\begin{eqnarray}\eqlabel{periodic-depth-uncert}
\frac{1}{{\sigma_\depth}^2} &=& \sum_{m=1}^{M(\period,\,\phase)}
    \frac{1}{{\delta\bar{\depth}_m}^2} \quad.
\end{eqnarray}
Note that this result has been \emph{marginalized} over the parameters of the
systematics model.
Therefore, this estimate of the uncertainty on the depth takes any
uncertainty that we have about the systematics into account.

In general, the variable depth model will \emph{always} get a higher
likelihood because it is more flexible.
Therefore, a formal model comparison is required to compete these two models
against each other on equal footing.
For computational simplicity and speed, we use the Bayesian Information
Criterion (BIC).
The traditional definition of the BIC is
\begin{eqnarray}
-\frac{1}{2}\,\BIC &=&
    \ln p(\flux_n\,|\,\period,\,\phase,\,\duration)
    - \frac{K}{2} \ln N
\end{eqnarray}
where the likelihood function is evaluated at the maximum, $K$ is an estimate
of the model complexity and $N$ is the effective sample size.
To emphasize that $K$ and $N$ are tuning parameters of the method, we
rewrite this equation as
\begin{eqnarray}
-\frac{1}{2}\,\BIC &=&
    \ln p(\flux_n\,|\,\period,\,\phase,\,\duration) -
        \frac{J\,\alpha}{2}
\end{eqnarray}
where $J$ is the number of allowed depths---one for the constant depth model
and the number of transits for the variable depth model---and $\alpha$ is
chosen heuristically.
For the K2 Campaign~1 dataset, we find that $\alpha \sim 1240$ leads to
reliable recovery of injected signals while still being fairly insensitive to
false signals.

To limit memory consumption, in the periodic search, we profile (or maximize)
over \phase\ and \duration\ subject to the constraint that
$\BIC_\mathrm{const} < \BIC_\mathrm{var}$ and requiring that the signal have
at least two observed transits.
This yields a one-dimensional spectrum of the signal-to-noise of the depth
measurement as a function of period using Equations~(\ref{eq:periodic-depth})
and (\ref{eq:periodic-depth-uncert}) to compute $\depth/\sigma_\depth$ at
each period.
The result is a generalization of the \project{BLS} frequency spectrum
\citep{Kovacs:2002} to a light curve model that includes both a transit and
the trends.
For example, \fig{periodic} shows the spectrum for a planet candidate
transiting EPIC 201613023.

After selecting the best candidate based on the signal-to-noise of the depth,
we mask out the sections of the linear search corresponding to these transits
and iterate the periodic search.
This permits us to find second transiting planets in light curves in which
we have already found a more prominent signal.
Under our assumption of independent transits, this masking procedure is
equivalent to removing the sections of data that have a transit caused by the
exoplanet that produces the highest peak.
For the purposes of this \paper, we iterate the periodic search until we find
three peaks for each light curve.
This will necessarily miss the smallest and longest period planets in systems
with more than three transiting planets but given the conservative vetting in
the next \sectionname, three peaks are sufficient to discover all the high
signal-to-noise transits.

\begin{figure}[p]
\begin{center}
\includegraphics{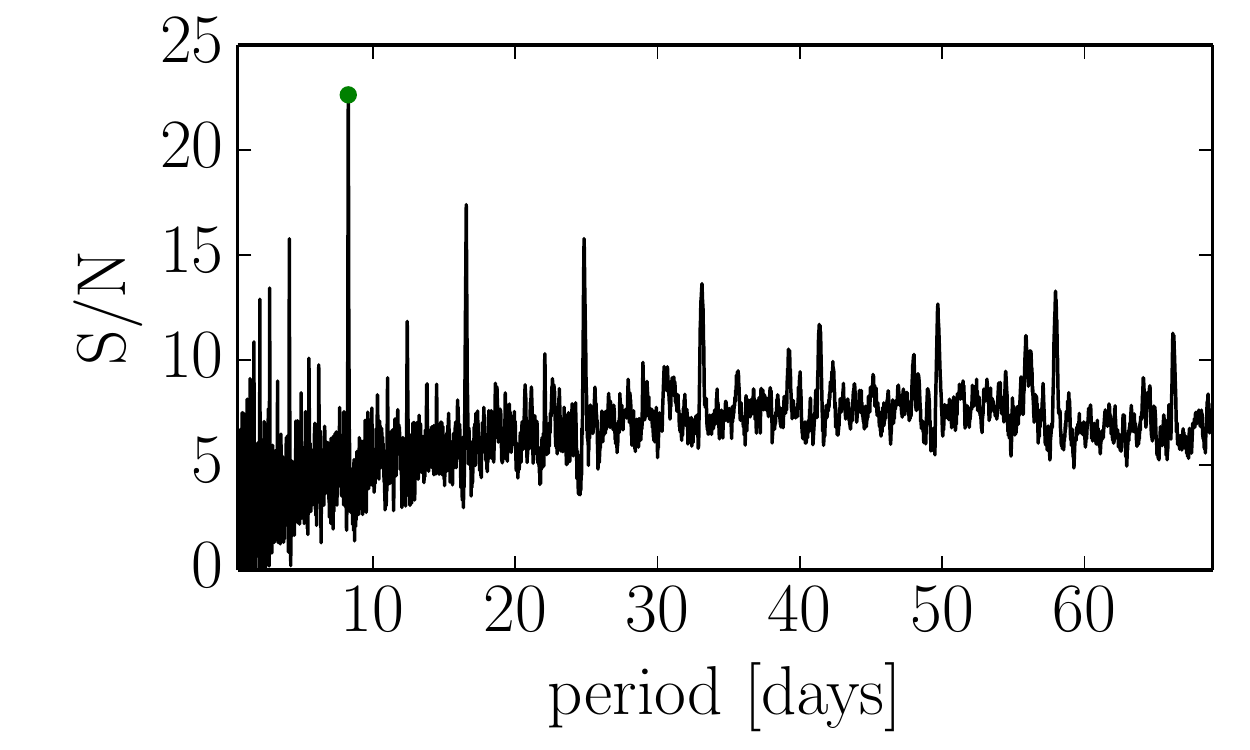}
\end{center}
\caption{%
The signal-to-noise spectrum as a function of period for the light curve of
EPIC 201613023.
This is the generalization of the \project{BLS} spectrum \citep{Kovacs:2002}
to this simultaneous model of the transit and the systematic trends.
To compute this spectrum, the results of the linear search (\fig{linear})
were used as described in \sect{search}.
The top peak (at a period of 8.3~days) is indicated with a green dot.
Iterating the periodic search found no other transit signals above the
signal-to-noise threshold.
\figlabel{periodic}}
\end{figure}

\paragraph{Initial candidate list}

The periodic search procedure returned three signals per target so this gave
an initial list of 65,109 candidates.
The vast majority of these signals are not induced by a transiting planet:
there are many false positives.
Therefore to reduce the search space, we estimate the signal-to-noise of each
candidate by comparing the peak height to a robust estimate of variance in
\BIC\ values across period.
This is not the same criterion used to select the initial three peaks but we
find that it produces a more complete and pure sample.
A cut in this quantity can reject most variable stars and low signal-to-noise
candidates that can't be reliably recovered from the data.
To minimize contamination from false alarms but maximize our sensitivity, we
choose a threshold of 15.
\response{In absolute value, this threshold is somewhat higher than the
standard signal-to-noise threshold used when searching for transits in the
\kepler\ light curves \citep[for example][]{Petigura:2013} but given the
larger amplitude of the systematic noise in the \KT\ light curves, it is not
surprising that a higher threshold is required to produce a manageable list of
candidates for hand vetting.
That being said, it is likely that a reduction in this threshold would yield
more discoveries at the cost of a larger set of hand classifications.}

We also find that the signals with periods $\lesssim 4$ days are strongly
contaminated by false alarms.
This might be because of the fact that our independence assumption
(\eqalt{indtran}) breaks down at these short periods.
Therefore, we discard all signals with periods shorter than 4 days,
acknowledging this will cause us to miss some planets
\citep{Sanchis-Ojeda:2014}.
After these cuts, 741 candidates remain; we examine these signals by hand.
The full list of peaks and their relevant meta data is available online
at\footnote{\datareleaseurl}.

\paragraph{Hand vetting}

After our initial cuts on the candidate list, the majority of signals are
still false alarms mostly due to variable stars or single outlying data
points.
It should be possible to construct a more robust machine vetting algorithm
that discards these samples without missing real transits but for the purposes
of this \paper, we simply inspect the light curve for each of the 741
candidates \emph{by hand} to discard signals that are not convincing transits.
The results of this vetting can be seen online\footnote{\datareleaseurl}.

Although de-trended light curves are never used in the automated analysis of
the data, when conditioned on a specific set of transit parameters, the model
produces an estimate of what the light curve would look like in the absence
of systematic effects.
This prediction is one of the plots that we examine when vetting candidates
by hand.
For example, \fig{de-trended} shows the maximum likelihood light curve for
EPIC 201613023 evaluated at the candidate period, phase, duration, and depth.
Similarly, \fig{folded} shows the same prediction folded on the 8.3~day
period of this candidate.

After visually inspecting 741 signals, 101 candidate transits pass and are
selected as astrophysical events.
Many of these signals are due to ``false positives'' such as eclipsing binary
systems, either as the target star or as a background ``blend.'' We address
this effect in the following \sectionname, where we separate the list of
candidates into a list of astrophysical false positives and planet candidates.

\begin{figure}[p]
\begin{center}
\includegraphics{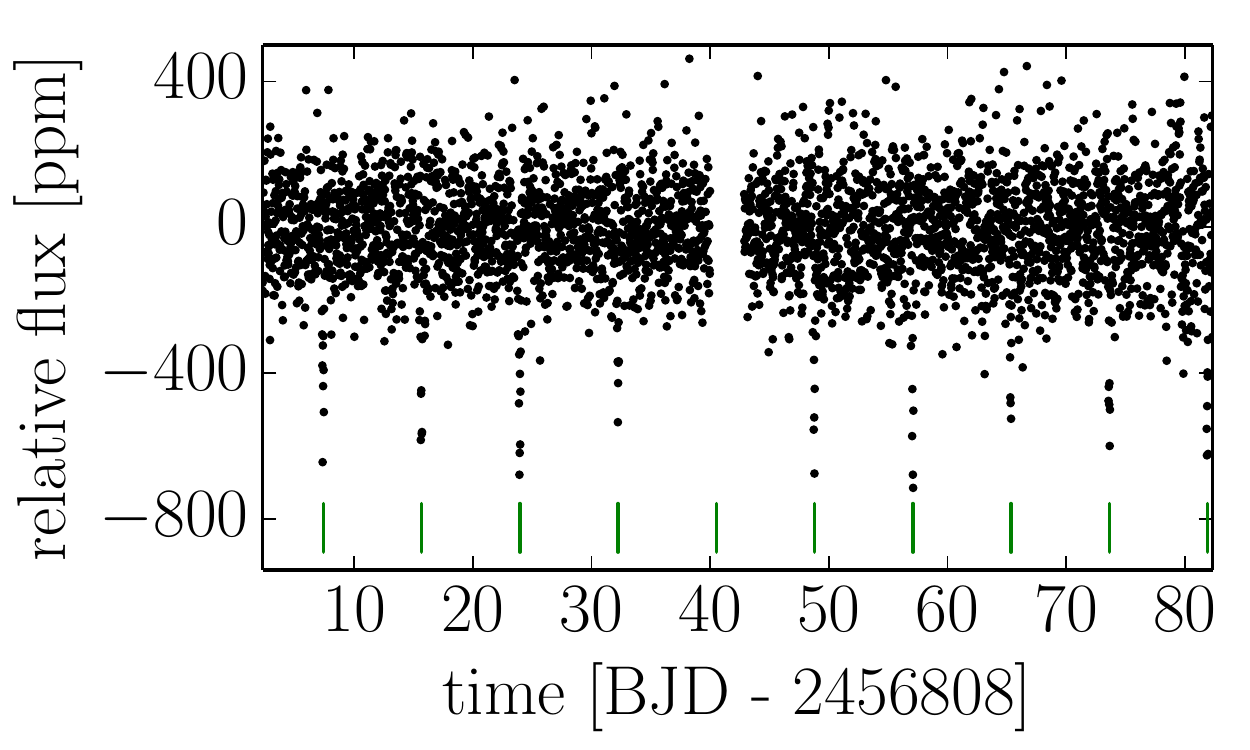}
\end{center}
\caption{%
The maximum likelihood ``de-trended'' light curve for EPIC 201613023
evaluated at the planet candidate's period, phase, duration, and depth.
The transit times are indicated by the green ticks below the light curve.
This Figure is only generated for qualitative hand vetting and in the search
procedure, the model is always marginalized over any choices about the
systematic trends.
\figlabel{de-trended}}
\end{figure}

\begin{figure}[p]
\begin{center}
\includegraphics{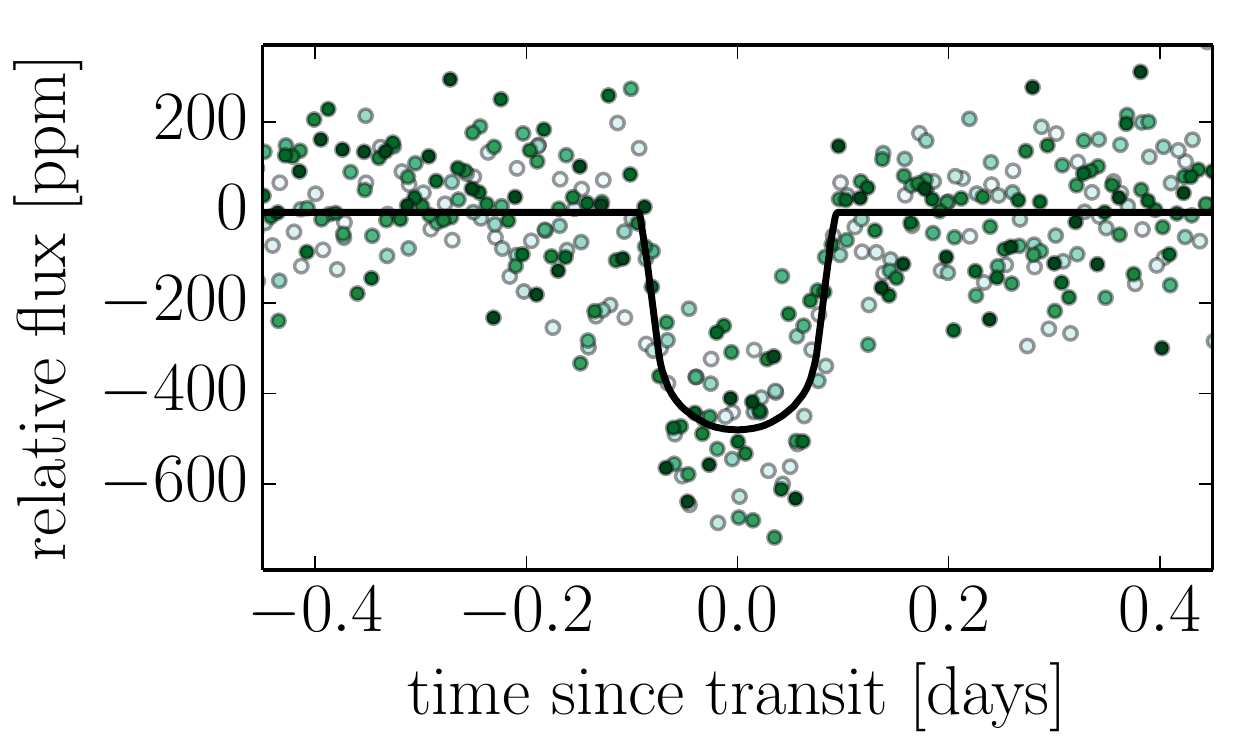}
\end{center}
\caption{%
The maximum likelihood prediction for the light curve of EPIC 201613023 (see
also \fig{de-trended}) folded on the 8.3~day period of this planet candidate.
The points are color-coded by time and the median \emph{a posteriori} transit
model is overplotted as a black line.
\figlabel{folded}}
\end{figure}

\paragraph{Astrophysical false positives}

A major problem with any transit search is the potential confusion
between transiting planets and stellar eclipsing binaries (EBs).  Of
particular concern are grazing stellar eclipses or stellar eclipses
that contribute only a small fraction of the total light in a
photometric aperture, resulting in greatly diluted eclipse depths able
to mimic the signals of small planets.

Ground-based transit surveys have experienced false-positive rates
well over 50~percent.  For example, \citet{Latham:2009} reported eight
eclipsing binaries and one transiting planet among the sample of
transit candidates in one field of the Hungarian Automated Telescope
Network transit search.  In fact, the follow-up process to try to rule
out such astrophysical false positives is a large portion of the
effort that goes into a transit survey
\citep[e.g.,][]{Odonovan:2006,Almenara:2009,Poleski:2010}.

Despite this large fraction of astrophysical false positives in
ground-based surveys, the primary \kepler\ Mission saw a much lower
false positive rate of only 5-10\% \citep{Morton:2011, Fressin:2013},
primarily due to three major factors.  First, the superior precision
of the \kepler\ photometry enables detection of secondary stellar
eclipses, odd-even transit depth variations, and ellipsoidal
variations \citep{Batalha:2010} to a much lower level than
ground-based surveys. Second, the relatively small pixels and stable
pointing of the \kepler\ telescope has enabled the identification of
many spatially distinct blended eclipsing binaries by means of
detailed pixel-level analysis \citep{Bryson:2013} to identify shifts
in the center of light during transits.  And finally, \kepler\ is
sensitive to much smaller planets than ground-based surveys, and small
planets are much more common than the Jupiter-sized planets able to be
detected from the ground.
\response{We note that while \citet{Santerne:2012} reported a $\sim$35\%
observational false positive rate, that study was exclusively focused on
short-period, large candidates, among which false positives are expected to be
more likely.
\citet{Desert:2015} has observationally confirmed a low false positive rate
for the majority of \kepler\ candidate parameter space.}

In \KT, the precision of the photometric tests used to vet for such
false positives  is lower and they must be applied
with care.  There are typically only a handful of transits, meaning
differences between ``odd'' and ``even'' transits must be large to
create a significant difference. Searching for ellipsoidal variations
is hindered by the short time baseline and the increased photometric
uncertainty in \KT\ data.  Centroid variations are feasible in
\KT\ but must be treated differently than in the original
\kepler\ mission where this effect was generally measured using
difference imaging \citep{Batalha:2010, Bryson:2013}.

To do first-pass vetting for blended EBs among our catalog of
planetary candidates, we test for significant centroid offsets using
the machinery that we have already established for modeling the
systematic trends in the data, inspired by the methods used to vet
\kepler\ candidates \citep{Bryson:2013}.
If any of the candidates have substantial centroid offsets in phase with their
transits, this indicates that the signal is likely caused by a background or
foreground transit of an eclipsing binary and we, therefore, remove it from
the final candidate list.
This is only an initial vetting step and a more complete characterization of
our catalog's reliability is forthcoming (Montet, \etal\ in preparation).

To measure \emph{centroid offsets}, we start by empirically measuring the
pixel centroid time series for each candidate by modeling the pixels near the
peak as a two-dimensional quadratic and finding the maximum at each time.
This method has been shown to produce higher precision centroid measurements
than center-of-light estimates (Vakili \etal, in preparation).
\Fig{centroid} shows the measured $x$ and $y$ pixel coordinate traces for EPIC
201613023.
Much like the photometry, this signal is dominated by the rigid body motion
of the spacecraft and we can, in fact, model it identically.
In our analysis, we model the light curve as a linear
combination of ELCs and a simple box transit model at a given period, phase,
and duration (\eqalt{linear-model}).
Under this model, the maximum likelihood depth can be computed analytically.
If we apply \emph{exactly the same model} to the centroid trace, the ``depth''
that we compute becomes the centroid motion in transit in units of pixels.
Since the motions won't necessarily point in a consistent direction across
transits, we treat each transit independently and report the average offset
amplitude weighted by the precision of each measurement.
To compute the significance of a centroid offset, we bootstrap the offset
amplitude for models at the same period and duration but randomly oriented
phases.
If the centroid measured for the candidate transit is substantially larger
than the random realizations, we label the candidate as a false positive.
In practice, the precision of the centroid measurements isn't sufficient to
robustly reject many candidates, but two candidates---EPIC 201202105 and EPIC
201632708---have offsets 3-$\sigma$ above the median out-of-transit offset
amplitude so they are removed from the final catalog.
For example, \Fig{offsets} shows the in-transit centroid offset measured for
EPIC 201202105 and compares it to the distribution of out-of-transit offset
amplitudes.

A quick \emph{a priori} estimate of the background blended eclipsing
binary rate serves as a good sanity check.  A query to the TRILEGAL
\citep[TRIdimensional modeL of thE GALaxy,][]{Girardi:2005} galaxy
line-of-sight simulation software reveals that the typical density of
field stars along the line of sight to the Campaign 1 field is about
$7.8\times 10^{-4}$\,arcsec$^{-2}$.  This gives a probability of about
0.16 that a background star might be blended within a 8\,arcsec radius
(two pixels) from a target star.  Allowing that $\sim$10\% of stars
might host close binary companions within the period range accessible
by this survey, this gives a probability of 0.016 that a blended
binary star might be chance-aligned within two pixels of any given
target star.  Noting that the average number of planets per star with
periods less than 30 days is about 0.25 \citep{Fressin:2013}, we can
roughly estimate that we expect $<$10\% of our candidates to be caused
by nearby contaminating EBs.  This estimate suggests that such
astrophysical false positives should be rare in our sample, consistent
with our detection of only 2 candidates with clear centroid offsets.

\begin{figure}[p]
\begin{center}
\includegraphics[width=0.4\textwidth]{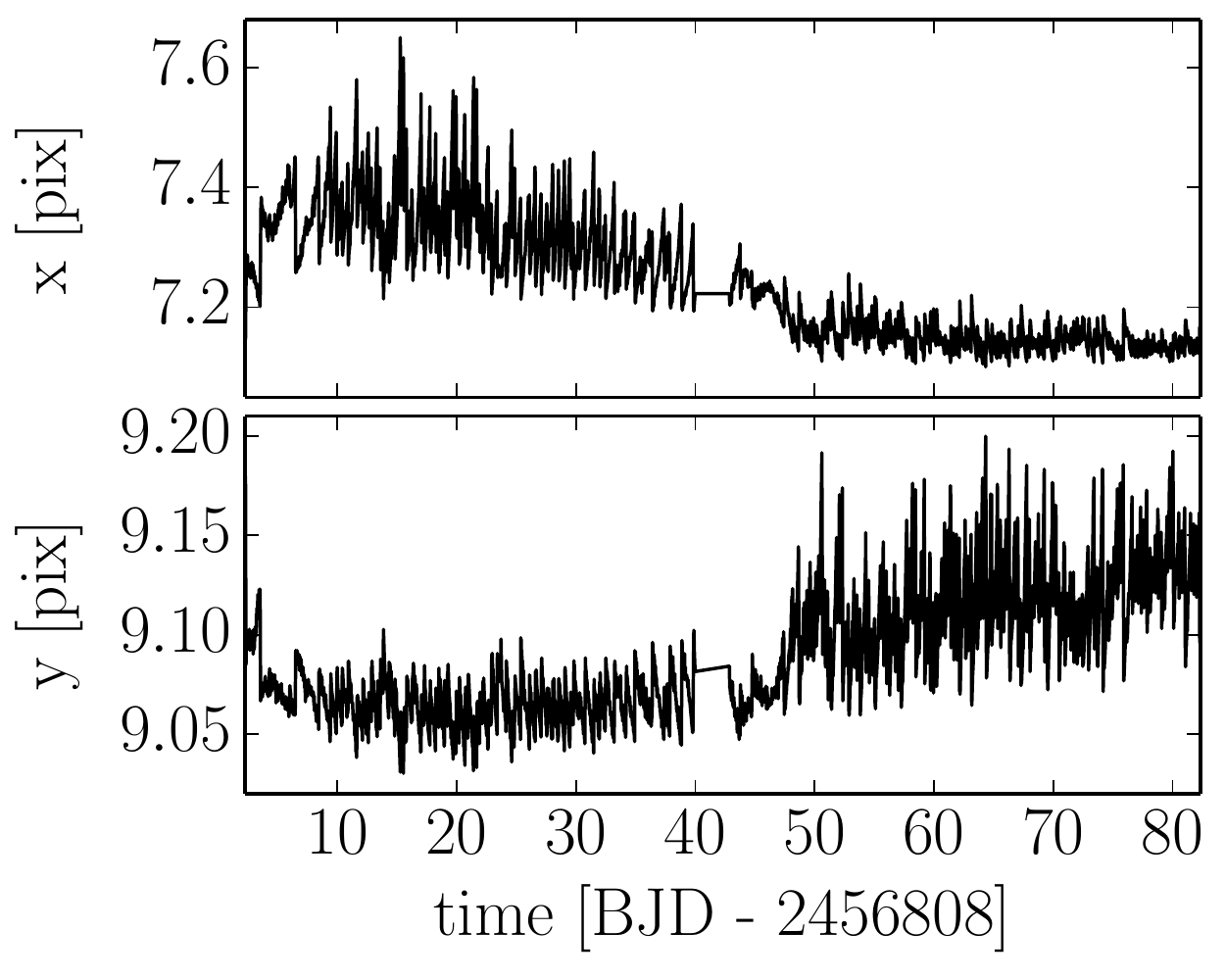}
\includegraphics[width=0.4\textwidth]{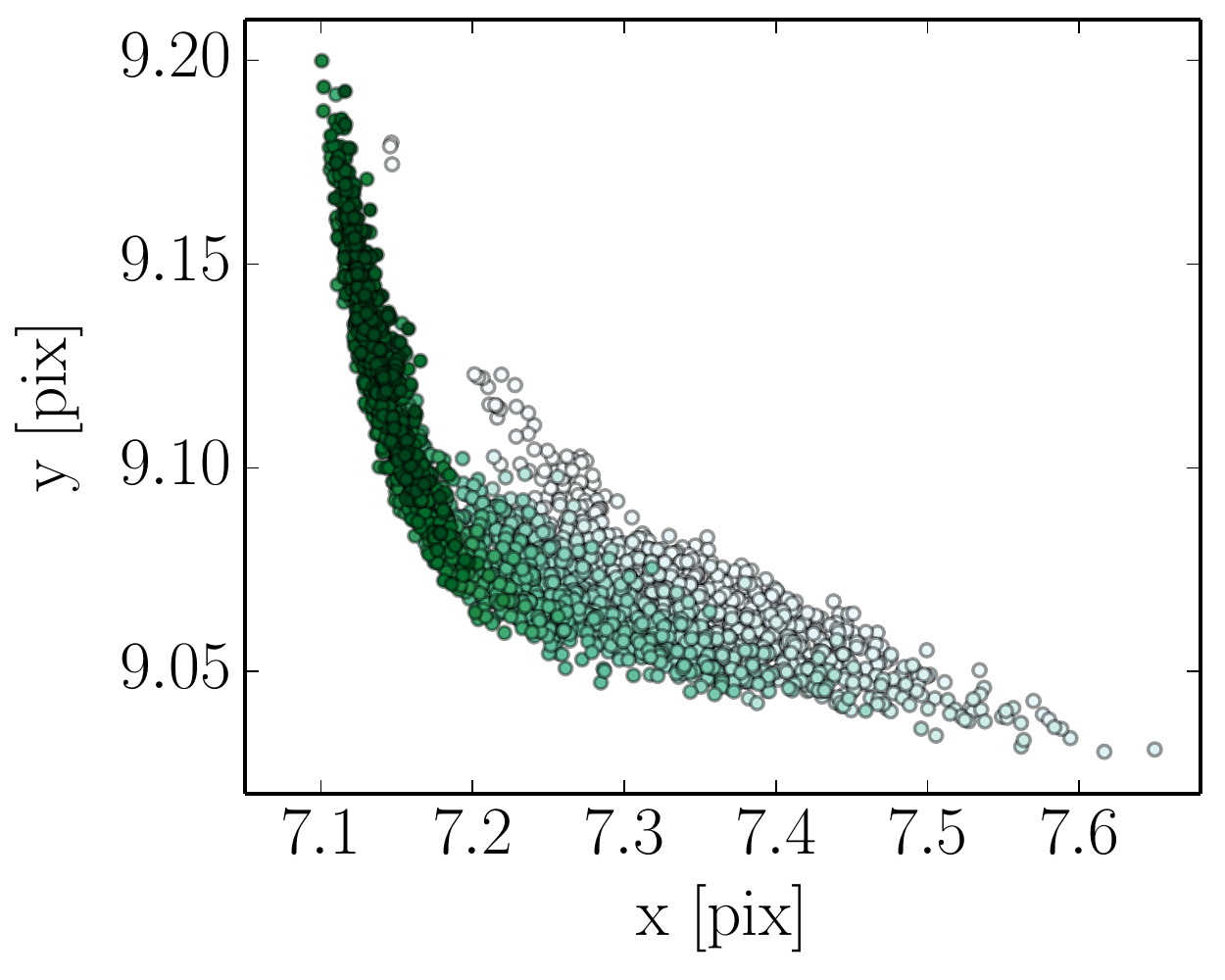}
\end{center}
\caption{%
The centroid motion for EPIC 201613023.
\emph{Left:} The measured $x$ and $y$ pixel coordinates as a function of time.
\emph{Right:} The pixel coordinates color-coded by time.
As identified by \citet{Vanderburg:2014}, the centroid motions fall in a
slowly time variable locus.
If the centroid coordinates in transit are inconsistent with the
out-of-transit motions, the candidate is likely to be an astrophysical false
positive.
\figlabel{centroid}}
\end{figure}

\begin{figure}[p]
\begin{center}
\includegraphics{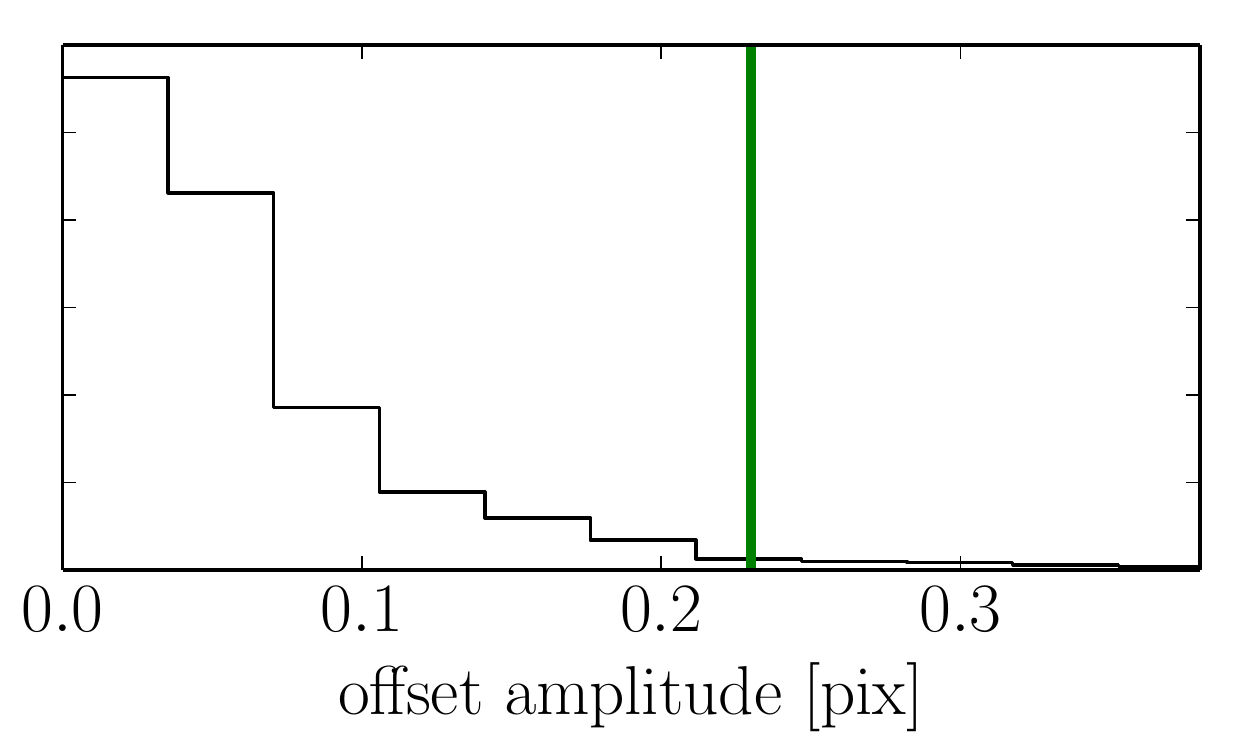}
\end{center}
\caption{%
The estimated in-transit centroid offset for EPIC 201202105 (green line)
compared to the distribution of 1000 centroid offests computed for randomly
assigned phases (black histogram).
The in-transit measurement is 3-$\sigma$ larger than the median out-of-transit
offset so it is rejected from the final catalog.
\figlabel{offsets}}
\end{figure}

\section{Performance}
\sectlabel{perform}

To test the performance and detection efficiency of our method, we conducted a
suite of injection and recovery tests, five per
star for all 21,703 target stars.
For each test, we inject the signal from a realistic planetary system into the
raw aperture photometry of a random target and run the resulting injected
light curve through the full pipeline (except the manual vetting).
If the search returns a planet candidate---passing all of the same cuts as we
apply in the main search (except the manual vetting)---with period and
reference transit time within 6~hours of the injected signal, we count that
injection as necovered.
The detection efficiency of the search is given approximately by the fraction
of recovered injections as a function of the relevant parameters.

To generate the synthetic signals, we use the following procedure:
\begin{enumerate}
{\item Draw the number of transiting planets based on the observed
multiplicity distribution of KOIs \citep{Burke:2014}.}
{\item Sample---from the distributions listed in
\tab{dist}---limb darkening parameters and, for each planet, an orbital
period, phase, radius ratio, impact parameter, eccentricity, and argument of
periapsis.}
{\item Based on the chosen physical parameters, simulate the light curve,
taking limb darkening and integration time into account \citep{Mandel:2002,
Kipping:2010}, and multiply it into the raw aperture photometry.}
\end{enumerate}
We then process these light curves using exactly the pipeline that we use for
the light curves without injections.
Finally, we test for recovery after applying the cuts in signal-to-noise and
period.
We should, of course, also vet the results of the injection tests by hand
to ensure that our measurements of detection efficiency aren't biased by the
hand-vetting
step but, since we chose to limit our sample to very high signal-to-noise
candidates, it seems unlikely that our hand vetting removed any true transit
signals.
Any estimates of the false alarm rate will, however, be affected by this
negligence but we leave a treatment of this for future work.

\figurename s~\figref{completeness} and~\figref{completeness-mag} show the
fraction of recovered signals as a function of the physical parameters of the
injection, and the magnitude of the star in the \kepler\ bandpass as reported
in the Ecliptic Plane Input Catalog
(EPIC\footnote{\url{http://archive.stsci.edu/k2/epic.pdf}}).
As expected, the shallower transits at longer periods are recovered less
robustly and all signals become harder to detect for fainter stars.
It is worth noting that these \figurename s are projections (or
marginalizations) of a higher dimensional measurement of the recovery rate as
a function of all of the input parameters.
For example, this detection efficiency map is conditioned on our assumptions
about the eccentricity distribution of planets and it is marginalized over the
empirical distribution of stellar parameters.
It is possible to relax this assumption and apply different distributions by
re-weighting the simulations used to generate this figure.
Therefore, alongside this \paper, we publish the full list of injection
simulations\footnote{\datareleaseurl} to be used for population inference
(occurrence rate measurements).

\response{%
While we argue that the most relevant quantity to use to quantify the
performance of a transit search pipeline is the efficiency with which it
discovers transits, it is also useful to consider some other standard metrics.
In particular, while de-trended light curves are never used at any stage of
the analysis, our method does make a prediction for the systematics model and
we can measure the relative precision of the residuals away from this model.
These residuals are what would be used as de-trended light curves if that was
the goal.
\Fig{performance} shows, as a function of the \kepler\ magnitude reported in
the EPIC, the 6-hour CDPP \citep{Christiansen:2012} for each light curve after
subtracting the best fit linear combination of 150 ELCs.
}

\begin{figure}[p]
\begin{center}
\includegraphics{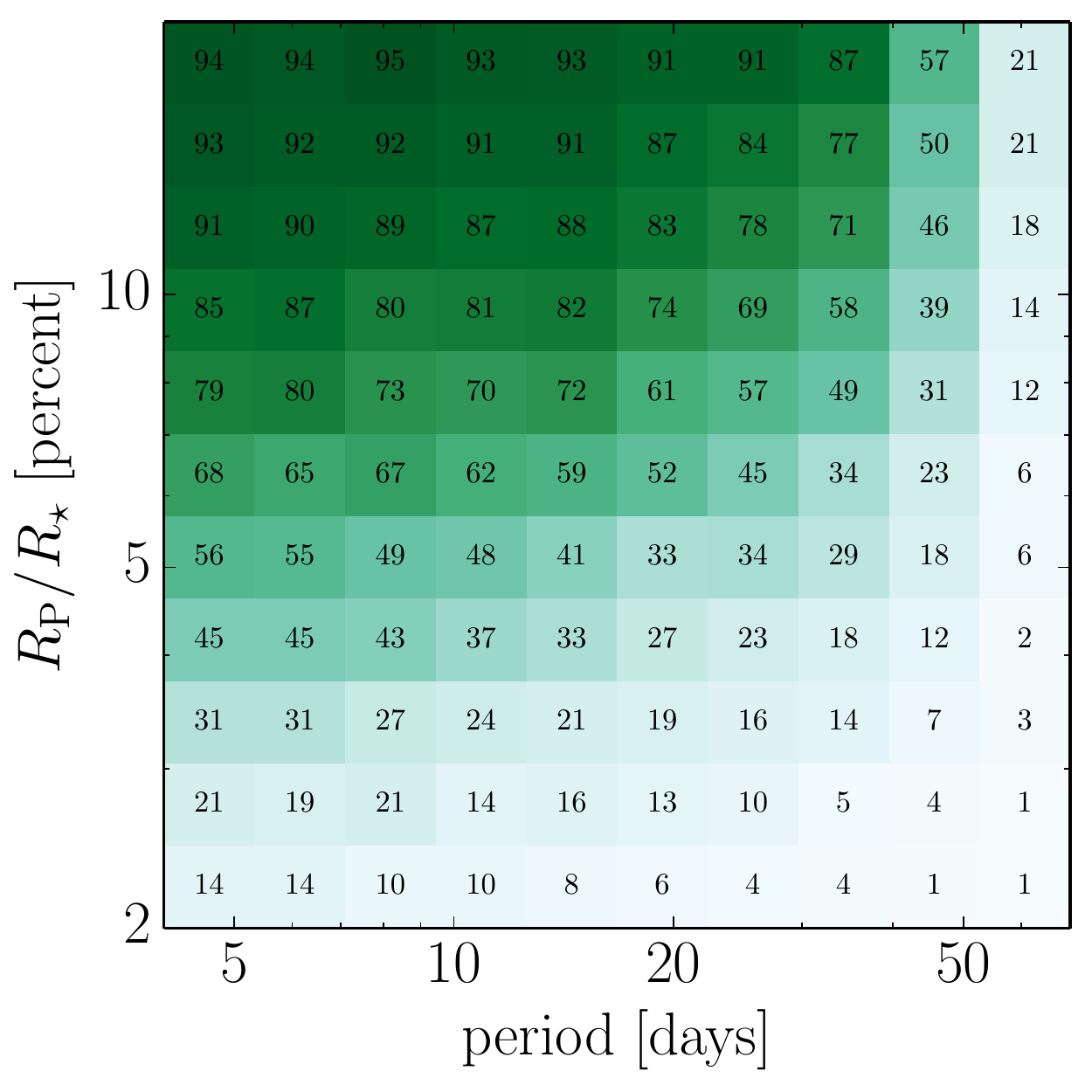}
\end{center}
\caption{%
The detection efficiency of the search procedure as a function of the
physical transit parameters computed empirically by
injecting synthetic transit signals into the raw light curves and measuring
the fraction that are successfully recovered.
These tests were performed on the entire set of stars so these numbers are
marginalized over all the stellar properties, including magnitude.
\figlabel{completeness}}
\end{figure}

\begin{figure}[p]
\begin{center}
\includegraphics{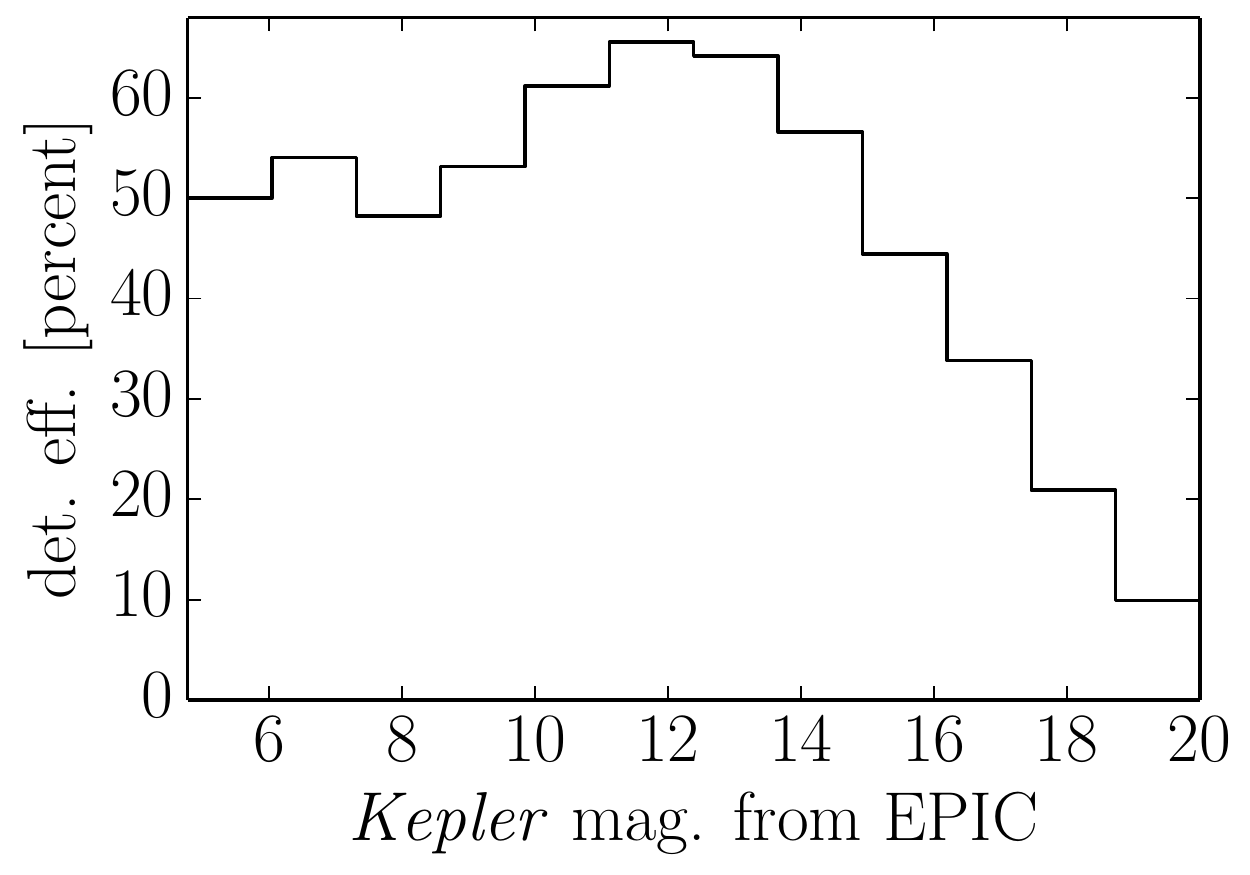}
\end{center}
\caption{%
Like \fig{completeness}, the empirically measured detection efficiency of the
search procedure as a function of stellar magnitude as reported by the
Ecliptic Plane Input Catalog.
This never reaches 90~percent because these numbers are marginalized over the
range of physical parameters shown in \fig{completeness}.
Even for the brightest stars, the long period, small transits cannot be
detected.
\figlabel{completeness-mag}}
\end{figure}

\begin{figure}[p]
\begin{center}
\includegraphics{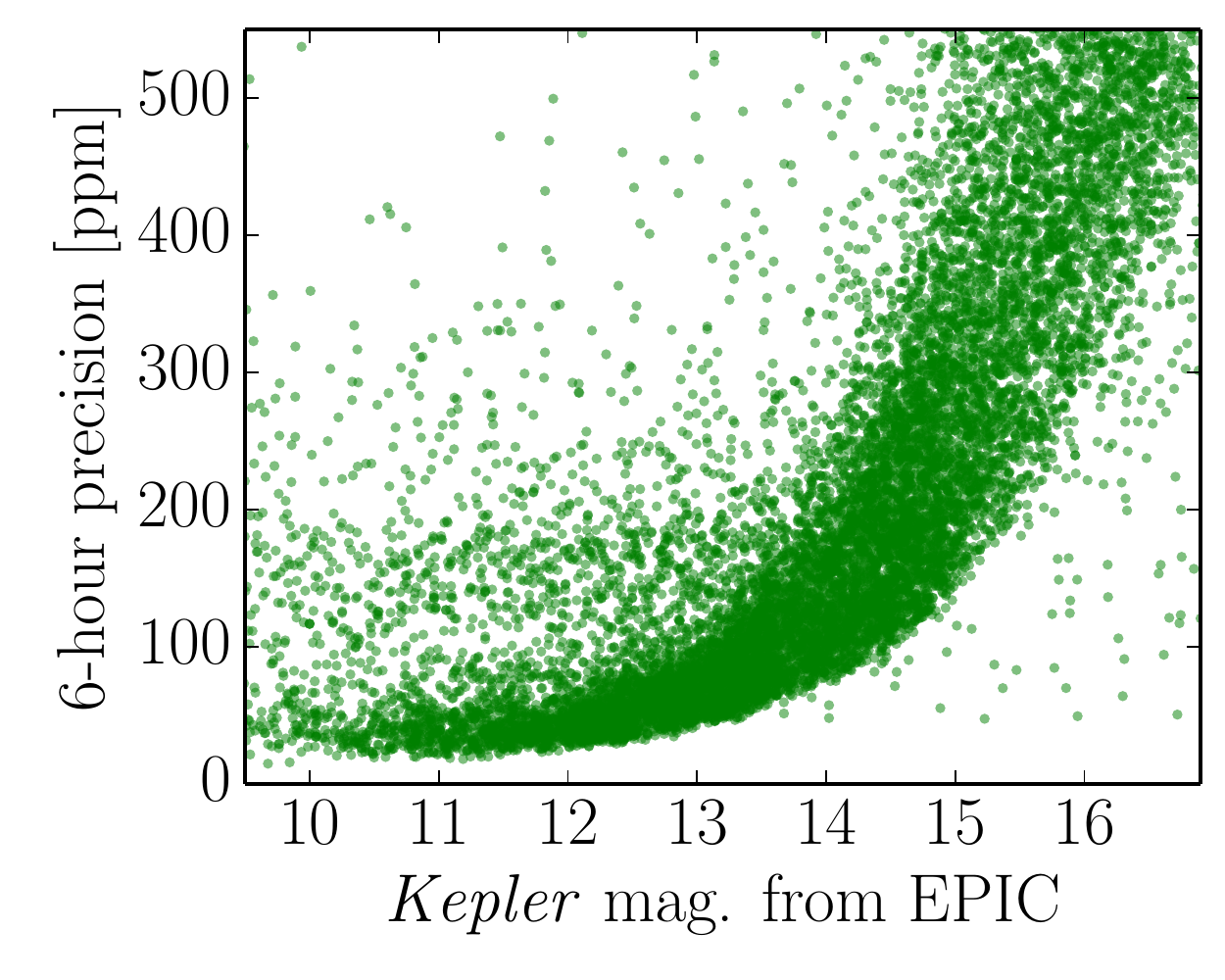}
\end{center}
\caption{%
\response{%
The 6-hour CDPP \citep{Christiansen:2012} for each light curve in Campaign 1
after subtracting the best fit linear combination of 150 ELCs.
For each star, the precision is plotted as a function of the \kepler\
magnitude reported in the Ecliptic Plane Input Catalog.
The ``outliers'' in the bottom right corner of the plot are caused by a
bright star within the photometric aperture and the points in the top left
corner of the plot are variable stars where the major trends in the light
curve are not caused by systematic effects, making the ELC model a bad fit.}
\figlabel{performance}}
\end{figure}

\section{Results}
\sectlabel{results}

Out of the 21,703 Campaign~1 light curves, our pipeline returns 741 signals
that pass the signal-to-noise and period cuts.
After hand vetting by the two first authors, this list is reduced to 101
convincing astrophysical transit candidates.
Of these, 36 signals---in 31 light curves---have no visible secondary
eclipse and are deemed planet candidates.
These planet candidates are listed in \tab{cand}.
The two candidates transiting EPIC 201367065 were previously published
\citep{Crossfield:2015} and the third planet in that system is found as the
third signal by our pipeline but it falls just below the signal-to-noise cut
so it is left out of the catalog for consistency.
This suggests that a less conservative cut in signal-to-noise and more
aggressive machine vetting could yield a much more complete catalog at smaller
radii and longer periods even with the existing dataset.

The remaining signals are caused by EBs with visible secondary eclipses.
In most cases, the search reports the secondary eclipse as a candidate and in
a few very high signal-to-noise cases, the period reported by the pipeline is
incorrect and multiple candidates correspond to the same transit.
It is important to note, however, that the choices made in the search were
heuristically tuned to find planets, not binaries, so our results are not
complete or exhaustive, especially at short orbital periods.
There are other methods specifically tuned to find EBs in \KT\
\citep[such as][]{Armstrong:2014, Armstrong:2015} and these catalogs contain
our full sample of EBs and more.

For the planet candidates, we perform a full physical transit fit to the light
curve.
To do this fit, we use Markov Chain Monte Carlo
\citep[MCMC;][]{Foreman-Mackey:2013} to
sample from the posterior probability for the stellar and planetary
parameters taking limb darkening and integration time into account.
In this fit, we continue to model the trends in the data as a linear
combination of the 150 ELCs but, at this point, we combine this with a
realistic light curve model \citep{Mandel:2002, Kipping:2013a}.
Even though we have no constraints on the stellar parameters, we also sample
over a large range in stellar mass and radius so that future measurements can
be applied by re-weighting the published samples.
In \tab{cand} we list the sample quantiles for the observable quantities and
the full chains are available electronically\footnote{\datareleaseurl}.
\fig{candidates} shows the observed distribution of planet candidates in the
catalog.

In a follow-up to this \paper, we will characterize the stars for each of the
candidates in detail but for now it's worth noting that many of the planet
candidates are orbiting stars selected for \KT\ as M-type stars.
If this rate remains robust after stellar characterization and if these
numbers are representative of the yields in upcoming \KT\ Campaigns, the \KT\
Mission will substantially increase the number of planets known to transit
cool stars.

\begin{figure}[p]
\begin{center}
\includegraphics{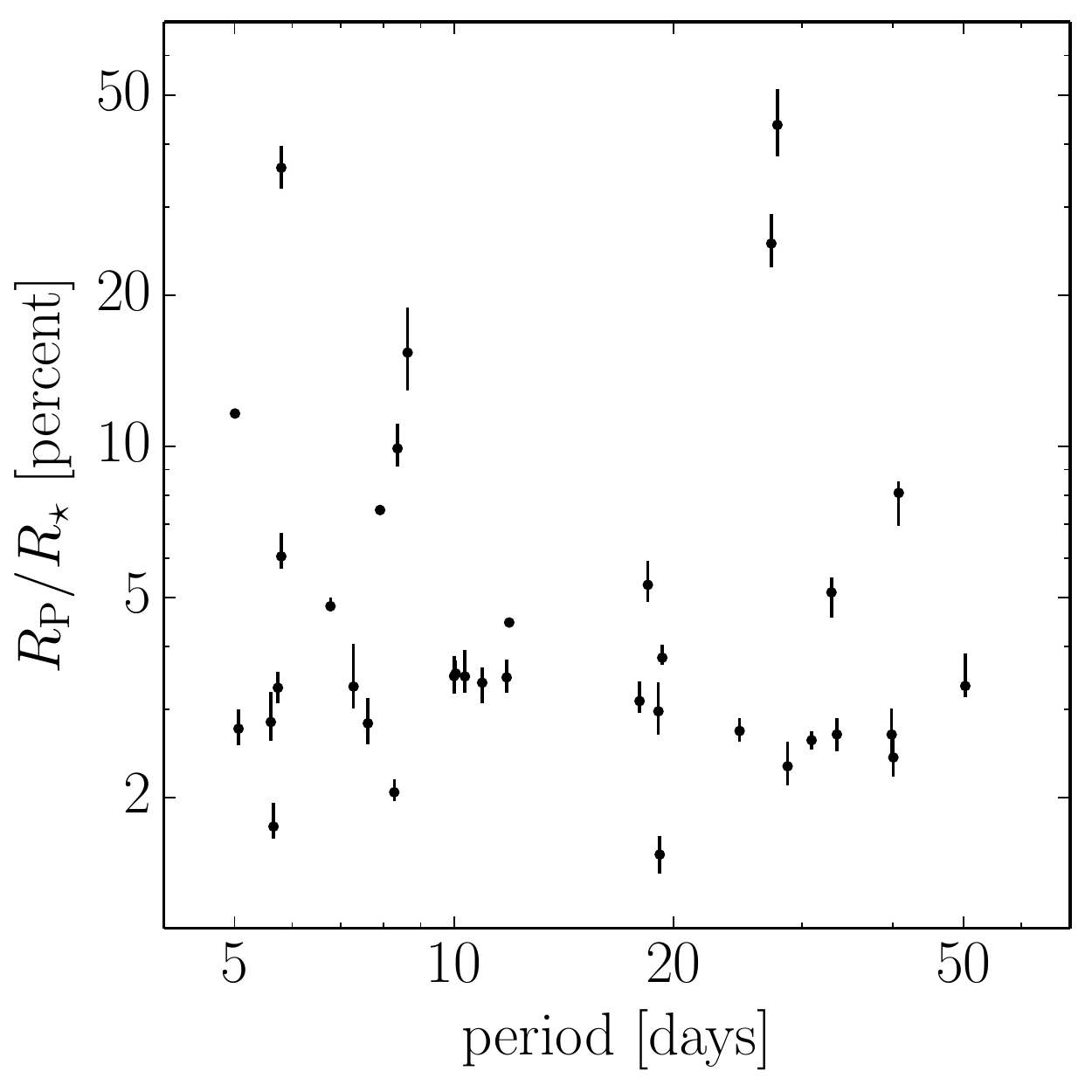}
\end{center}
\caption{%
The \emph{a posteriori} distribution of planet candidates in the catalog.
The error bars indicate the 0.16 and 0.84 posterior sample quantiles for the
radius ratios.
\figlabel{candidates}}
\end{figure}

\section{Discussion}

We have searched the \KT\ Campaign~1 data set for exoplanet transit signals.
Our search is novel because it includes a very flexible systematics model,
which is fit simultaneously with the exoplanet signals of interest (and
marginalized out).
By this method, we find 36 transiting exoplanets, which we have vetted by both
automatically and manually and characterized by probabilistic modeling.
The candidates are listed in \tab{cand} and posterior distributions of planet
candidate properties are available\footnote{\datareleaseurl}.

The flexible systematics model we employ is a 150-parameter linear combination
of PCA components derived from the full set of 21,703 stellar light curves.
That is, it presumes that the systematics afflicting each star are shared in
some way across other stars \response{while the astrophysical signals are
causally unconnected \citep{Scholkopf:2015}}.
\response{This assumption means that, while there is no formal guarantee that
the basis contains no astrophysical signals, it is unlikely that the top
components will be contaminated.}
It is our belief---although not a strict assumption of our model---that the
systematics are caused primarily by pointing drifts, or movements of the
pixels in the focal plane relative to the stars.
In principle, if the systematics \emph{are} dominated by pointing issues, the
systematics model could require only three parameters---three Euler
angles---not 150 amplitudes.
However, because (as the pointing drifts) each star sees its own unique
local patch of flat-field variations, the mapping from pointing drifts to
brightness variations can be extremely non-linear.
Furthermore, because when the pointing is moving fast there is a smearing of
the point-spread function, there are effects keyed to the time derivative of
the Euler angles as well.
The large number (150) of linear coefficients gives the linear model the
freedom to model complex non-linear behavior; we are trading off parsimony
in parameters with the enormous computational advantages of maintaining
linearity (and therefore also convexity).
The computational advantages of the linear model are three-fold:
Convexity obviates searching in parameter space for alternative modes;
linear least-squares optimization can be performed with simple linear algebra;
given Gaussian uncertainties and uninformative priors, marginalizations over
the linear parameters also reduces to pure linear algebra.

The goal of this \paper\ was to get exoplanet candidates out of the
\KT\ pixel-level data, it was \emph{not} to generate light curves.
That is, both the search phase and the characterization phase of the method
are approximations to computations of a likelihood function for the pixel
data telemetered down from the satellite.
We did not generate ``corrected'' or ``pre-search conditioned'' light-curves
at any stage; we simultaneously fit systematics and the signals of interest
to the raw data.
For this reason, there is no sense in which this method ever really
produces corrected light curves.
% For us, a systematics-corrected light curve is not really a thing.

In this work, we are agnostic about fundamental properties of the host stars.
The only assumptions we make are that the star targeted by the \KT\
team is truly the planet host, and that there is no dilution by other stars in
any aperture.
As a result, these posterior distributions reflect the maximum possible
uncertainty in parameters such as the planet radius, which depend sensitively
on properties of the host star.
To use these distributions to characterize the properties of specific systems,
one could re-weight our samples using a measurement of the inferred stellar
properties.

This project does not live in isolation and this is certainly not the last
time the \KT\ data will be searched!
There are other teams searching the \KT\ light curves for transiting planets
(A.~Vanderburg, private comm.) and they are likely to find some planets that
we did not and vice versa.
We make many heuristic choices and short-cuts in this search.
For example, the choice to work at 150 principal components was based on
computational feasibility and qualitative tests on a handful of light curves
instead of any real model selection or utility optimization.

Another major limitation is that, in principle, the systematics model is
designed to describe spacecraft-induced trends, but not intrinsic stellar
variability.
In practice, the method can still find planets around variable stars but a
more sophisticated model should be more robust in this case.
One appealing option would be to model the systematics as a Gaussian Process
where the input parameters are both time and the same 150 ELCs.
Interestingly, while this model isn't linear, the search and marginalization
can still be executed efficiently---using optimized linear algebra algorithms
\citep[][Foreman-Mackey et al.\ in preparation]{Ambikasaran:2014}---inside the
search loop.

Additionally, while we apply this systematics model simultaneously with a
transiting planet model to search for planet candidates, this scheme is
not restricted to planet searches.
Any astrophysical event that could be observed in the \KT\ data could be
searched for in the same way.
By modeling a set of ELCs with any arbitrary data model, events in the \KT\
data that appear similar to that data model could be identified.
Such a technique may be useful in searching for astrophysical events such as
ellipsoidal variations induced by orbiting companions,
stellar activity,
microlensing events, especially in the upcoming Campaign 9,
or active galactic nuclei variability.

A substantial caveat to the reliability of all existing transiting exoplanet
searches is that they all include human intervention.
This makes quantifying the false alarm rate of these catalogs complicated.
There has been some work on automated vetting algorithms using supervised
classification algorithms \citep{McCauliff:2014, Jenkins:2014} but these
methods rely on hand classified examples for training and the performance is
not yet competitive with human classification.

The catalog of planet candidates presented here includes only planets with
periods longer than 4 days and at least two transits in the \KT\ Campaign~1
footprint.
This means that we are necessarily missing many planets with orbital periods
outside  this range.
In particular, planets with a single transit in the dataset must be
abundant.
These candidates are the most relevant for the study of planetary system
formation and for statistical inference of the distribution of habitable zone
exoplanets.
What's more, given the observing strategy for \tess, where each field will
only be contiguously observed for one month at a time, methods for finding and
characterizing planets with a single transit are vital and the new \KT\ light
curves are a perfect test bed.

As a supplement to this \paper, we make all the results, data products, and
MCMC chains available at \datareleaseurl.
The \LaTeX\ source for this \paper, complete with the full revision history,
is available at \url{http://github.com/dfm/k2-paper} and the pipeline
implementation is available at \url{http://github.com/dfm/ketu} under the MIT
open-source software license.
This code and a lot of computation time are all that is needed to reproduce
the \figurename s in this \paper.

\acknowledgments
It is a pleasure to thank
Eric Agol (UW),
Ruth Angus (Oxford),
Tom Barclay (Ames),
Zach Berta-Thompson (MIT),
Daniel Bramich (QEERI, Qatar),
G\'eza Kov\'acs (Konkoly Observatory),
Laura Kreidberg (Chicago),
Erik Petigura (Berkeley),
Roberto Sanchis Ojeda (Berkeley),
and
Andrew Vanderburg (Harvard)
for helpful contributions to the ideas and code presented here.
We also thank the anonymous referee for comments that improved the
manuscript.
DFM, DWH, and DW were partially supported by the National Science Foundation
(grant IIS-1124794),
the National Aeronautics and Space Administration
(grant NNX12AI50G), and the Moore--Sloan Data Science Environment at NYU.
BTM was supported by a National Science Foundation Graduate Research
Fellowship (grant DGE‐1144469).

This research made use of the NASA \project{Astrophysics Data System} and the
NASA Exoplanet Archive.
The Archive is operated by the California Institute of Technology, under
contract with NASA under the Exoplanet Exploration Program.
This \paper\ includes data collected by the \kepler\ mission. Funding for the
\kepler\ mission is provided by the NASA Science Mission directorate.
We are grateful to the entire \kepler\ team, past and present.
Their tireless efforts were all essential to the tremendous success of the mission
and the successes of \KT, present and future.
These data were obtained from the Mikulski Archive for Space Telescopes
(MAST).
STScI is operated by the Association of Universities for Research in
Astronomy, Inc., under NASA contract NAS5-26555.
Support for MAST is provided by the NASA Office of Space Science via grant
NNX13AC07G and by other grants and contracts.

{\it Facilities:} \facility{Kepler}

\appendix

\section{Mathematical model}
\sectlabel{math}

We model the raw aperture photometry as a linear combination of 150 ELCs and
a transit model.
Formally, this can be written for the light curve of the $k$-th star as
\begin{eqnarray}\eqlabel{linear-model}
\bvec{f}_k &=& \bvec{A}\,\bvec{w}_k + \mathrm{noise}
\end{eqnarray}
where
\begin{eqnarray}
\bvec{f}_k &=& \left (\begin{array}{cccc}
    f_{k,1} & f_{k,2} & \cdots & f_{k,N}
\end{array}\right )^\T
\end{eqnarray}
is the list of aperture fluxes for star $k$ observed at $N$ times
\begin{eqnarray}
\bvec{t} &=& \left (\begin{array}{cccc}
    t_{1} & t_{2} & \cdots & t_{N}
\end{array}\right )^\T \quad.
\end{eqnarray}
In \eq{linear-model}, the design matrix is given by
\begin{eqnarray}
\bvec{A} &=& \left (\begin{array}{cccccc}
    x_{1,1} & x_{2,1} & \cdots & x_{150,1} & 1 & m_\bvec{\theta}(t_1) \\
    x_{1,2} & x_{2,2} & \cdots & x_{150,2} & 1 & m_\bvec{\theta}(t_2) \\
    && \vdots &&&\\
    x_{1,N} & x_{2,N} & \cdots & x_{150,N} & 1 & m_\bvec{\theta}(t_N)
\end{array}\right )
\end{eqnarray}
where the $x_{j,n}$ are the basis ELCs---with the index $j$ running over
components and the index $n$ running over time---and $m_\bvec{\theta}(t)$ is
the transit model
\begin{eqnarray}
m_\bvec{\theta}(t) &=& \left\{\begin{array}{cl}
-1 & \mathrm{if\,}t\,\mathrm{in\,transit} \\
0 & \mathrm{otherwise}
\end{array}\right.
\end{eqnarray}
parameterized by a period, phase, and transit duration (these parameters are
denoted by \bvec{\theta}).

Assuming that the uncertainties on $\bvec{f}_k$ are Gaussian and constant,
the maximum likelihood solution for \bvec{w} is
\begin{eqnarray}
{\bvec{w}_k}^* &\gets& \left( \bvec{A}^\T\,\bvec{A} \right)^{-1}\,
                       \bvec{A}^\T\,\bvec{f}_k
\end{eqnarray}
and the marginalized likelihood function for the transit depth is a Gaussian
with the mean given by the last element of ${\bvec{w}_k}^*$ and the variance
given by the lower-right element of the matrix
\begin{eqnarray}
{\bvec{\delta w}_k}^2 &\gets& {\sigma_k}^2 \,
            \left( \bvec{A}^\T\,\bvec{A} \right)^{-1}
\end{eqnarray}
where $\sigma_k$ is the uncertainty on $\bvec{f}_k$.
The amplitude of this Gaussian is given by
\begin{eqnarray}\eqlabel{depth-likelihood}
\mathcal{L}_k &=& \frac{1}{(2\,\pi\,{\sigma_k}^2)^{N/2}}\,\exp\left(
-\frac{1}{2\,{\sigma_k}^2}\,
\left| \bvec{f}_k - \bvec{A}\,{\bvec{w}_k}^* \right|^2
\right)
\end{eqnarray}
evaluated at the maximum likelihood value ${\bvec{w}_k}^*$.

\clearpage
\bibliography{k2}
\clearpage

\clearpage

\begin{table}[p]
\begin{center}
\begin{tabular}{lcc}
\toprule
Parameter & Units & Distribution \\
\midrule

limb darkening parameters $q_1$ and $q_2$ & --- & $q \sim U(0,\,1)$ \\
orbital period \period & days & $\ln \period \sim U(\ln 0.5,\,\ln 70)$ \\
reference transit time \phase & days & $\phase \sim U(0,\,\period)$ \\
radius ratio $R_P/R_\star$ & --- & $\ln R_P/R_\star \sim U(\ln 0.02,\,\ln 0.2)$ \\
impact parameter \impact & --- & $\impact \sim U(0,\,1)$ \\
eccentricity \ecc & --- & $\ecc \sim \mathrm{Beta}(0.867,\,3.03)$ \\
argument of periapsis \pomega & --- & $\pomega \sim U(-\pi,\,\pi)$ \\

\bottomrule
\end{tabular}
\end{center}
\caption{%
The distribution of physical parameters for the injected signals.
The eccentricity distribution is based on \citet{Kipping:2013} and the
limb darkening parameterization is given by \citet{Kipping:2013a}.
\tablabel{dist}}
\end{table}

\begin{table}[p]
\begin{center}
\footnotesize
\begin{tabular}{ccccccc}
\toprule
EPIC & \kepler\ mag & RA (J2000) & Dec (J2000) & \period\ [days] & $t_0$ [BJD-2456808] & $R_\mathrm{P} / R_\star$ \\
\midrule
201208431 &14.41& 174.745640 & -3.905585 & $10.0040_{-0.0016}^{+0.0018}$ & $7.5216_{-0.0090}^{+0.0098}$ & $0.0349_{-0.0026}^{+0.0034}$ \\
201257461 &11.51& 178.161109 & -3.094936 & $50.2677_{-0.0074}^{+0.0083}$ & $20.3735_{-0.0098}^{+0.0147}$ & $0.0334_{-0.0017}^{+0.0054}$ \\
201295312 &12.13& 174.011630 & -2.520881 & $5.6562_{-0.0007}^{+0.0007}$ & $3.7228_{-0.0091}^{+0.0086}$ & $0.0175_{-0.0009}^{+0.0020}$ \\
201338508 &14.36& 169.303502 & -1.877976 & $10.9328_{-0.0021}^{+0.0022}$ & $6.5967_{-0.0081}^{+0.0088}$ & $0.0339_{-0.0030}^{+0.0025}$ \\
201338508 &14.36& 169.303502 & -1.877976 & $5.7350_{-0.0006}^{+0.0006}$ & $0.8626_{-0.0055}^{+0.0054}$ & $0.0331_{-0.0023}^{+0.0025}$ \\
201367065 &11.57& 172.334949 & -1.454787 & $10.0542_{-0.0004}^{+0.0004}$ & $5.4186_{-0.0018}^{+0.0018}$ & $0.0354_{-0.0011}^{+0.0022}$ \\
201367065 &11.57& 172.334949 & -1.454787 & $24.6470_{-0.0016}^{+0.0014}$ & $4.2769_{-0.0029}^{+0.0030}$ & $0.0272_{-0.0013}^{+0.0016}$ \\
201384232 &12.51& 178.192260 & -1.198477 & $30.9375_{-0.0052}^{+0.0029}$ & $19.5035_{-0.0039}^{+0.0053}$ & $0.0260_{-0.0011}^{+0.0011}$ \\
201393098 &13.05& 167.093771 & -1.065755 & $28.6793_{-0.0116}^{+0.0105}$ & $16.6212_{-0.0177}^{+0.0305}$ & $0.0231_{-0.0020}^{+0.0028}$ \\
201403446 &11.99& 174.266344 & -0.907261 & $19.1535_{-0.0050}^{+0.0050}$ & $7.3437_{-0.0143}^{+0.0116}$ & $0.0154_{-0.0013}^{+0.0014}$ \\
201445392 &14.38& 169.793665 & -0.284375 & $10.3527_{-0.0011}^{+0.0011}$ & $5.6110_{-0.0051}^{+0.0047}$ & $0.0349_{-0.0025}^{+0.0045}$ \\
201445392 &14.38& 169.793665 & -0.284375 & $5.0644_{-0.0006}^{+0.0006}$ & $5.0690_{-0.0064}^{+0.0059}$ & $0.0274_{-0.0020}^{+0.0025}$ \\
201465501 &14.96& 176.264468 & 0.005301 & $18.4488_{-0.0015}^{+0.0015}$ & $14.6719_{-0.0032}^{+0.0035}$ & $0.0531_{-0.0039}^{+0.0061}$ \\
201505350 &12.81& 174.960319 & 0.603575 & $11.9069_{-0.0004}^{+0.0005}$ & $9.2764_{-0.0015}^{+0.0013}$ & $0.0446_{-0.0006}^{+0.0009}$ \\
201505350 &12.81& 174.960319 & 0.603575 & $7.9193_{-0.0001}^{+0.0001}$ & $5.3840_{-0.0008}^{+0.0006}$ & $0.0747_{-0.0013}^{+0.0016}$ \\
201546283 &12.43& 171.515165 & 1.230738 & $6.7713_{-0.0001}^{+0.0001}$ & $4.8453_{-0.0011}^{+0.0012}$ & $0.0481_{-0.0012}^{+0.0020}$ \\
201549860 &13.92& 170.103081 & 1.285956 & $5.6083_{-0.0006}^{+0.0005}$ & $4.1195_{-0.0047}^{+0.0045}$ & $0.0283_{-0.0023}^{+0.0041}$ \\
201555883 &15.06& 176.075940 & 1.375947 & $5.7966_{-0.0002}^{+0.0002}$ & $5.3173_{-0.0050}^{+0.0027}$ & $0.0604_{-0.0032}^{+0.0068}$ \\
201565013 &16.91& 176.992193 & 1.510249 & $8.6381_{-0.0002}^{+0.0003}$ & $3.4283_{-0.0015}^{+0.0016}$ & $0.1538_{-0.0243}^{+0.0355}$ \\
201569483 &11.77& 167.171299 & 1.577513 & $5.7969_{-0.0000}^{+0.0000}$ & $5.3130_{-0.0003}^{+0.0002}$ & $0.3587_{-0.0334}^{+0.0379}$ \\
201577035 &12.30& 172.121957 & 1.690636 & $19.3062_{-0.0013}^{+0.0013}$ & $11.5790_{-0.0027}^{+0.0025}$ & $0.0380_{-0.0012}^{+0.0023}$ \\
201596316 &13.15& 169.042002 & 1.986840 & $39.8415_{-0.0155}^{+0.0136}$ & $21.8572_{-0.0101}^{+0.0120}$ & $0.0267_{-0.0022}^{+0.0034}$ \\
201613023 &12.14& 173.192036 & 2.244884 & $8.2818_{-0.0007}^{+0.0006}$ & $7.3752_{-0.0052}^{+0.0055}$ & $0.0205_{-0.0008}^{+0.0012}$ \\
201617985 &14.11& 179.491659 & 2.321476 & $7.2823_{-0.0008}^{+0.0007}$ & $4.6337_{-0.0050}^{+0.0050}$ & $0.0333_{-0.0032}^{+0.0072}$ \\
201629650 &12.73& 170.155528 & 2.502696 & $40.0492_{-0.0259}^{+0.0186}$ & $4.5363_{-0.0172}^{+0.0202}$ & $0.0241_{-0.0020}^{+0.0025}$ \\
201635569 &15.55& 178.057026 & 2.594245 & $8.3681_{-0.0002}^{+0.0002}$ & $3.4514_{-0.0014}^{+0.0015}$ & $0.0991_{-0.0078}^{+0.0120}$ \\
201649426 &13.22& 177.234262 & 2.807619 & $27.7704_{-0.0001}^{+0.0001}$ & $13.3476_{-0.0002}^{+0.0001}$ & $0.4365_{-0.0583}^{+0.0777}$ \\
201702477 &14.43& 175.240794 & 3.681584 & $40.7365_{-0.0025}^{+0.0026}$ & $3.5451_{-0.0025}^{+0.0026}$ & $0.0808_{-0.0114}^{+0.0043}$ \\
201736247 &14.40& 178.110797 & 4.254747 & $11.8106_{-0.0019}^{+0.0016}$ & $3.8483_{-0.0071}^{+0.0093}$ & $0.0347_{-0.0024}^{+0.0030}$ \\
201754305 &14.30& 175.097258 & 4.557340 & $19.0726_{-0.0049}^{+0.0048}$ & $1.4893_{-0.0133}^{+0.0128}$ & $0.0297_{-0.0030}^{+0.0042}$ \\
201754305 &14.30& 175.097258 & 4.557340 & $7.6202_{-0.0011}^{+0.0012}$ & $3.6813_{-0.0057}^{+0.0061}$ & $0.0281_{-0.0026}^{+0.0034}$ \\
201779067 &11.12& 168.542699 & 4.988131 & $27.2429_{-0.0001}^{+0.0001}$ & $12.2599_{-0.0003}^{+0.0002}$ & $0.2535_{-0.0259}^{+0.0369}$ \\
201828749 &11.56& 175.654342 & 5.894323 & $33.5093_{-0.0018}^{+0.0023}$ & $5.1554_{-0.0032}^{+0.0037}$ & $0.0267_{-0.0020}^{+0.0021}$ \\
201855371 &13.00& 178.329775 & 6.412261 & $17.9715_{-0.0017}^{+0.0015}$ & $9.9412_{-0.0038}^{+0.0033}$ & $0.0311_{-0.0017}^{+0.0030}$ \\
201912552 &12.47& 172.560460 & 7.588391 & $32.9410_{-0.0032}^{+0.0039}$ & $28.1834_{-0.0105}^{+0.0057}$ & $0.0513_{-0.0056}^{+0.0035}$ \\
201929294 &12.97& 174.656969 & 7.959611 & $5.0084_{-0.0001}^{+0.0001}$ & $4.5703_{-0.0012}^{+0.0022}$ & $0.1163_{-0.0014}^{+0.0011}$ \\
\bottomrule
\end{tabular}

\end{center}
\caption{%
The catalog of planet candidates and their observable properties.
These values and their uncertainties are derived from MCMC samplings and the
numbers are computed as the 0.16, 0.5, and 0.84 posterior sample quantiles.
The coordinates are retrieved directly from the EPIC.
\tablabel{cand}}
\end{table}

\end{document}